\documentclass[aps,prd, nofootinbib,
preprintnumbers,showkeys, reprint,
superscriptaddress]{revtex4-1}

\usepackage{graphicx}
\usepackage{tikz}
\usepackage{multirow}
\usetikzlibrary{shapes.misc}

\tikzset{
    cross/.pic = {
    \draw[rotate = 45, very thick] (-#1,0) -- (#1,0);
    \draw[rotate = 45, very thick] (0,-#1) -- (0, #1);
    }
}
\usepackage{physics}
\usetikzlibrary{decorations.markings}
\usepackage{subcaption}
\usepackage{dcolumn}
\usepackage{bm}
\usepackage{amsmath, amssymb}
\usepackage{wasysym}
\usepackage{float}
\usepackage{multirow}
\usepackage{xcolor}
\usepackage{scrextend}
\usepackage{booktabs, array}
\usepackage[colorlinks=true, 
pdfstartview=FitV, 
bookmarks=true, 
bookmarksnumbered=true, 
breaklinks]{hyperref}
\usepackage{color}
\usepackage[normalem]{ulem} 
\definecolor{blue}{rgb}{0.0, 0.0, 1.0}
\definecolor{red}{rgb}{1.0, 0.0, 0.0}
\definecolor{royalblue}{rgb}{0.0, 0.14, 0.4}
\definecolor{purple}{rgb}{75.0, 0.0, 230.0}
\definecolor{teal}{rgb}{0.0, 128.0, 128.0}
\hypersetup{linkcolor=blue, 
	citecolor=purple, 
	urlcolor=blue}

\def\orcid#1{\kern .08em\href{https://orcid.org/#1}{\includegraphics[keepaspectratio,width=0.7em]{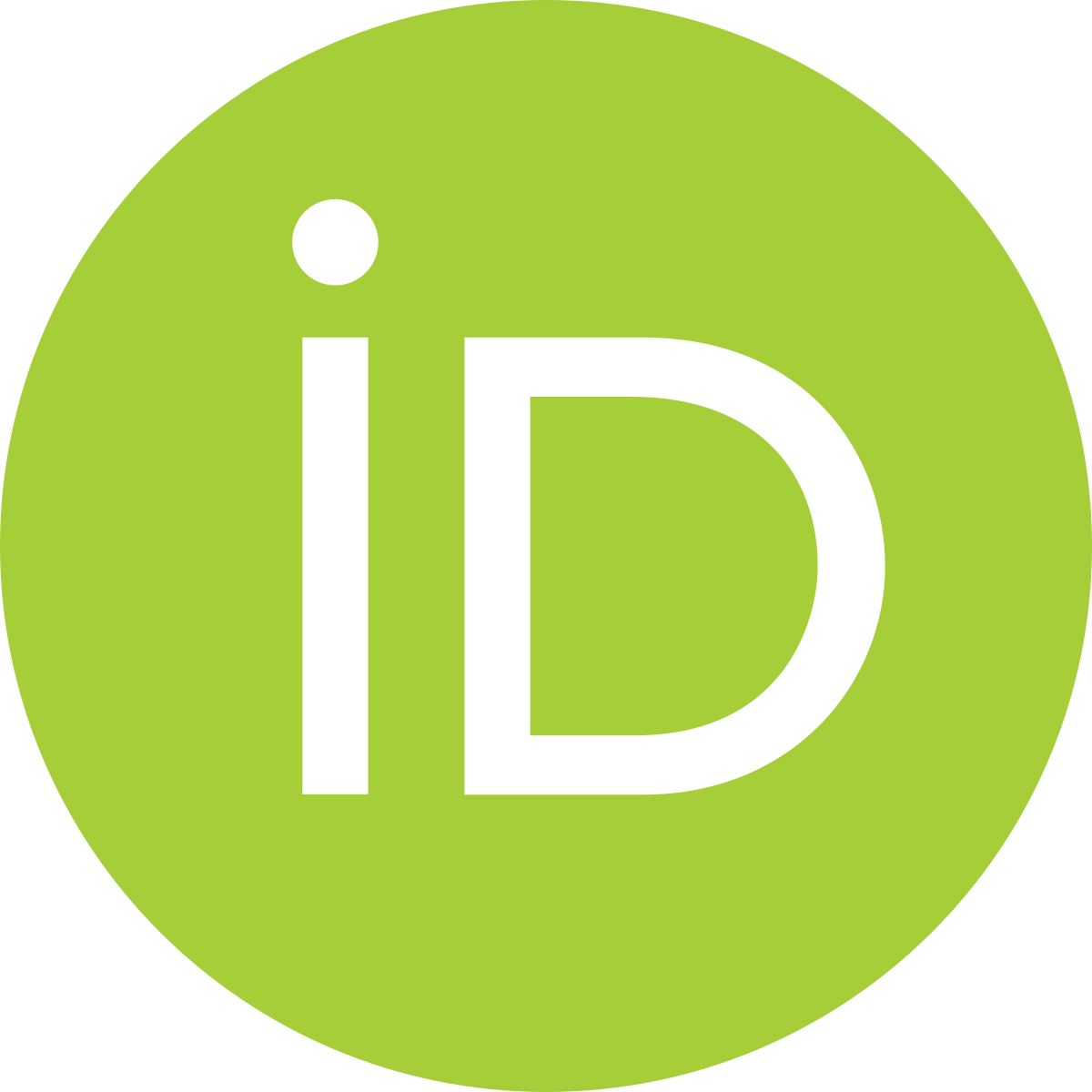}}}

\definecolor{ttcolor}{RGB}{240,248,255}
\definecolor{btcolor}{RGB}{144,238,144}
\definecolor{tbcolor}{RGB}{216,191,216}
\definecolor{bbcolor}{RGB}{255,228,225}

\usepackage{soul}
\usepackage{booktabs}

\usepackage{xcolor}
\usepackage{tikz}
\usepackage{neuralnetwork}
\usepackage{etoolbox}
\usepackage{listofitems}
\usetikzlibrary{decorations.pathreplacing}
\usetikzlibrary{fadings}

\begin{document}
\title{Revisiting the $\Lambda(1405)$ pole structure with convolutional neural networks}

\author{Julius B. Pagayon\orcid{0009-0009-2800-2570}}
\email[]{jbpagayon@up.edu.ph}
\affiliation{National Institute of Physics, University of the Philippines Diliman, Quezon City 1101, Philippines}

\author{Vince Angelo A. Chavez\orcid{0009-0009-2373-1985}}
\email[]{vachavez@up.edu.ph}
\affiliation{National Institute of Physics, University of the Philippines Diliman, Quezon City 1101, Philippines}

\author{Denny Lane B. Sombillo\orcid{0000-0001-9357-7236}}
\email[]{dbsombillo@up.edu.ph}
\affiliation{National Institute of Physics, University of the Philippines Diliman, Quezon City 1101, Philippines}

\date{\today}
\begin{abstract}
    \noindent
    We revisit the long-standing ambiguity surrounding the complex pole structure of the $\Lambda(1405)$ resonance by reframing it as a classification problem for convolutional neural networks (CNNs). By training our models to recognize subtle geometric variations on nearly degenerate lineshapes using targeted differential feature on empirical CLAS data, we establish a data-driven consensus on large inference samples. Our results show that the analytic structure characterized by two poles on the $[bt]$ sheet and additional pole on the $[bb]$ sheet globally dominates. The inference-guided pole parameter extraction reveals that the $\Lambda(1405)$ is a two-state system, characterized by a molecular state sitting below the $\bar{K}N$ threshold and a non-molecular state lying above the $\Sigma \pi$ threshold.
\end{abstract}

\keywords{uniformization, deep learning, exotic hadrons}


\maketitle


\section{Introduction}\label{sec:intro} 

The analytic structure of the scattering matrix plays a central role in understanding the properties of hadronic systems. Owing to the requirements of analyticity, unitarity, and causality, the singularities of the $\mathcal{S}$-matrix contain essential information regarding the physical states participating in a scattering process. In particular, unstable states are associated with poles located on unphysical Riemann sheets of the complex energy plane, while branch points emerge whenever new reaction channels become kinematically accessible. The positions of these poles determine the rest masses and decay widths of resonances, whereas their residues characterize the coupling strengths to the corresponding channels. Consequently, the extraction and interpretation of $\mathcal{S}$-matrix poles have become one of the principal tools for investigating the internal structure of hadrons \cite{Eden:1966dnq,Taylor}.

Experimentally, these singularities manifest themselves as enhancements in scattering cross-sections and invariant-mass distributions. Nevertheless, the relationship between an experimentally observed lineshape and its underlying pole structure is often far from straightforward \cite{Olsen:2017bmm,Guo:2017jvc}. While some enhancements correspond to compact resonances, others originate from hadronic molecules generated through coupled-channel interactions \cite{Hanhart:2014ssa,Hyodo:2014bda}. In addition, purely kinematical effects may also produce resonance-like structure such as cusping effects arising from branch-point singularities associated with the opening of new channels \cite{Guo:2019twa,Liu:2015taa}. Because threshold effects, coupled-channel dynamics, and nearby singularities may all contribute simultaneously to the observed lineshape, distinguishing among these mechanisms remains one of the central challenges of hadron spectroscopy \cite{Guo:2014iya}.

From a wider perspective, hadron spectroscopy delivers an important avenue for exploring the nonperturbative regime of quantum chromodynamics (QCD). Since the strong coupling constant becomes large at low energies, perturbative techniques are no longer applicable, demanding alternative approaches for investigating the properties of strongly interacting particles. One such approach relies on the experimental determination of the hadronic spectrum, from which the organization of quarks and gluons within hadrons can be inferred \cite{JPAC:2021rxu}. The increasing number of observed states has revealed that the hadronic spectrum extends far beyond the predictions of the traditional quark model, giving rise to a wide variety of exotic configurations\cite{Brambilla:2019esw,Esposito:2016noz}, including compact multiquark states \cite{Yang:2020atz,LHCb:2019kea,Ali:2017wsf}; hadronic molecules \cite{Meng:2022ozq,Guo:2017jvc}, and hybrid states \cite{Woss:2020ayi,Meyer:2015eta}.

Among these exotic hadron candidates, the $\Lambda(1405)$ occupies a unique position. First observed in the early 1960s through kaon-induced reactions, the $\Lambda(1405)$ has long attracted attention because of its unusual properties \cite{Dalitz:1959dn,Dalitz:1960du}. Although its quantum numbers are consistent with those of a conventional strange baryon, its mass lies very close to the $\bar{K}N$ threshold, suggesting that meson-baryon interactions play an essential role in its formation. This observation led Dalitz and Tuan to propose that the $\Lambda(1405)$ could arise from coupled-channel interactions involving the $\bar{K}N$ and $\Sigma\pi$ channels rather than from a simple three-quark configuration. Consequently, the $\Lambda(1405)$ has become one of the most extensively studied resonances in hadron spectroscopy and a benchmark system to understand the nature of exotic hadrons \cite{Hyodo:2011ur,Mai:2020ltx}.

Over the past decades, increasingly precise experimental measurements have substantially improved our knowledge of the $\Lambda(1405)$. Among these, the measurements performed by the CLAS Collaboration have played a particularly important role. Through the reaction $\gamma p \rightarrow K^{+}\Sigma\pi$, the collaboration reconstructed the invariant-mass distributions of the $\Sigma\pi$ system and observed a distinct enhancement around $1405~\mathrm{MeV}$ \cite{CLAS:2013rjt}. These observations suggest that the underlying structure of the $\Lambda(1405)$ is strongly influenced by threshold effects and coupled-channel dynamics, indicating that the resonance cannot be adequately described by a simple Breit–Wigner parametrization. On top of it, the measured lineshapes exhibited significant differences among the various charge channels.

The $\Lambda(1405)$ is perhaps the most prominent example of a resonance whose nature remains strongly debated. Among the most influential frameworks are the chiral unitary approaches, which combine effective field theory with the requirements of unitarity and analyticity. In these models, the $\Lambda(1405)$ is not introduced as an explicit degree of freedom but instead emerges dynamically through the interaction of coupled meson-baryon channels, including
$\bar{K}N, \Sigma\pi, \eta\Sigma$, and $K\Xi$. Early works by Kaiser, Siegel, and Weise \cite{Kaiser:1995eg}, Oset and Ramos \cite{Oset:1997it}, and Oller and Meissner \cite{Oller:2000fj} demonstrated the importance of coupled-channel dynamics in generating the resonance. Later investigations by Jido \textit{et al.} \cite{Jido:2003cb} revealed the existence of two nearby poles associated with the observed enhancement, giving rise to the now well-known two-pole nature of the $\Lambda(1405)$ \cite{Meissner:2020khl}. This picture has since been supported by more sophisticated analyses including constraints from kaonic hydrogen measurements, photoproduction experiments, and higher-order corrections within chiral perturbation theory \cite{Sadasivan:2018jig, Kamano:2011ih, Mai:2012dt, Feijoo:2018den, Cieply:2016jby, Guo:2012vv, Ikeda:2012au, Ikeda:2011pi}.

Beyond chiral unitary approaches, the $\Lambda(1405)$ has also been investigated using dynamical coupled-channel models and potential models. Dynamical coupled-channel models explicitly account for the interplay among multiple scattering channels and have highlighted the essential role of threshold effects in shaping the observed line spectra \cite{Anisovich:2020lec, Anisovich:2019exw, Fernandez-Ramirez:2015tfa, Zhuang:2024udv, Xie:2023jve}. In contrast, potential models describe the interaction through effective meson-baryon potentials and often interpret the $\Lambda(1405)$ as a hadronic molecule, a compact three-quark state, or a superposition of different configurations, depending on the assumptions employed \cite{Marri:2021ymy, Revai:2019ipq, Shevchenko:2011ce}. More recently, lattice QCD calculations have provided additional insight into the role of meson-baryon interactions as well as the possible molecular nature of the $\Lambda(1405)$, although extracting the complete pole structure remains a considerable challenge \cite{Hall:2014uca}.

Despite decades of investigation, no consensus has yet been reached regarding the true nature of the $\Lambda(1405)$. Chiral unitary approaches generally favor a two-pole interpretation, whereas other analyses have obtained different conclusions depending on the treatment of the scattering amplitude and the assumptions imposed on the model. As emphasized in the comprehensive review by Mai \cite{Mai:2020ltx}, the persistence of these discrepancies suggests that complementary approaches are needed to reduce the influence of model-dependent assumptions. More recently, the BESIII Collaboration reported evidence for a two-pole structure associated with the $\Lambda(1405)$. The extracted pole positions differ from those commonly obtained in chiral unitary approaches, providing an independent constraint on the possible pole structure of the resonance. The analysis employed a Flatte-type parametrization, within which the analytic continuation of the amplitude gives rise to poles on different Riemann sheets \cite{BESIII:2024jgy}. On the other hand, Yamada and Morimatsu employed the uniformized Mittag–Leffler expansion, a model-independent method based on the analytic structure of the $\mathcal{S}$-matrix, and demonstrated that existing experimental data may also be interpreted within a single-pole framework \cite{Yamada:2021cjo}. This realization has motivated the study of more robust data-driven methods, including machine learning techniques, as an alternative means of investigating the pole structure of the $\Lambda(1405)$.

Machine learning techniques have recently emerged as powerful tools for investigating these problems \cite{Alghamdi:2026psc,Aarts:2025gyp,He:2023zin,Boehnlein:2021eym}. From this perspective, the interpretation of lineshapes can be formulated as a classification problem in which the input corresponds to an experimentally observed distribution and the output corresponds to the physical mechanism responsible for the observed enhancement. Once the classification scheme has been established, a deep neural network (DNN) can be trained to learn the mapping between these two spaces. The effectiveness of this procedure stems from the ability of neural networks to approximate highly nonlinear functions \cite{Hornik:1989yye,Schmidhuber:2014bpo}. The first application of this methodology to hadronic systems was reported in Ref. \cite{Sombillo:2020ccg}, where a neural network was developed to discriminate between bound-state and virtual-state enhancements near the nucleon-nucleon threshold. This framework was subsequently extended to the pion-nucleon system, in which a DNN was used to identify pole configurations on different Riemann sheets \cite{Sombillo:2021rxv}. More recently, machine learning methods have been applied to the investigation of exotic hadrons \cite{Ng:2021ibr,Fernandez-Ramirez:2019koa,Santos:2024bqr, Co:2024bfl,Zhang:2023czx,Liu:2022uex,Chen:2022ddj,Pagayon:2026byc}.

Motivated by these developments, the present work investigates the pole structure of the $\Lambda(1405)$ using convolutional neural networks. Rather than constructing a phenomenological model of the meson-baryon interaction, the proposed framework aims to learn the relationship between resonance lineshapes and their corresponding pole structures directly from the data. By comparing the predictions of the trained network with the experimental lineshapes obtained by the CLAS Collaboration, we seek to provide an independent perspective on the longstanding debate regarding the nature of the $\Lambda(1405)$.

The remainder of this paper is organized as follows: in Sec.~\ref{sec:2}, we discuss the theoretical framework of complex pole structures and the nature of hadronic resonances in $\mathcal{S}$-matrix theory. In Sec.~\ref{sec:3}, we describe the machine learning framework used to identify the dominant pole structure class. Section~\ref{sec:4} presents the global pole classification results inferred by the CNN ensemble. In Sec.~\ref{sec:5}, we perform the inference-guided extraction of pole parameters. Finally, we conclude with a summary and outlook in Sec.~\ref{sec:6}.

\section{Pole structure and nature of hadronic resonance\label{sec:2}}

To unambiguously classify the pole structure of $\Lambda(1405)$, we utilize phenomenological models based on formal scattering theory. In scattering processes, the incoming and outgoing states are connected via a unitary scattering operator $\hat{\mathcal{S}}$ via~\cite{Taylor}
\begin{equation}
    \ket{\Psi_{\text{out}}} = \hat{\mathcal{S}} \ket{\Psi_{\text{in}}}.
\end{equation}
This operator commutes with the free Hamiltonian $\hat{H}_0$ and the total angular momentum operator $\hat{J}$; thus, it is diagonal in the partial-wave basis $\ket{E,l,m}$ reducing its effect to a set of scalar coefficients $\mathcal{S}_l(p)$, known as the partial-wave $\mathcal{S}$-matrix elements. For a spherically symmetric potential $V(r)$, the $\mathcal{S}_l(p)$ can be determined by solving the radial Schrödinger equation (S.E.)
\begin{equation}
    \left[ \frac{\mathrm{d}^2}{\mathrm{d}r^2} - \frac{l(l+1)}{r^2} + p^2 - \lambda U(r) \right]\Psi_{l,p} (r) = 0,
\end{equation}
where $\lambda$ is the coupling strength and $U(r) \equiv 2mV(r)$. The radial S.E. is characterized by the asymptotic forms of the \textit{normalized wave function}
\begin{equation}
    \psi_{l,p}(r) \xrightarrow{r \to \infty} \frac{i}{2} \left[ \hat{h}_l^-(pr) - \mathcal{S}_l (p) \hat{h}_l^+ (pr) \right],
\end{equation}
and the \textit{regular wave function}
\begin{equation}
    \varphi_{l,p}(r) \xrightarrow{r \to \infty} \frac{i}{2} \left[ \mathcal{Q}_l(p)\, \hat{h}_l^{-} (pr) -  \mathcal{Q}^*_l(p)\, \hat{h}_l^{+} (pr)\right],
\end{equation}
expressed in terms of the Ricatti-Hankel functions $\hat{h}^{\pm}(z)$. Here, the momentum-dependent coefficient $\mathcal{Q}_l(p)$ is the \textit{Jost function}~\cite{Taylor, Newton:1982}, defined integrally as
\begin{equation}
    \mathcal{Q}_l(p) = 1 + \frac{\lambda}{p} \int_0^\infty \mathrm{d}r'\, \hat{h}_l^+(pr') U(r') \varphi_{l,p}(r').
\end{equation}
It can be shown that $\varphi_{l,p}(r) = \mathcal{Q}_{l,p}(r) \psi_{l,p}(r)$ implying that the Jost function is simply a ratio of the regular solution with that of the normalized one. Finally, it can be demonstrated that $\mathcal{S}_l(p)$ is related to $\mathcal{Q}_l(p)$ via
\begin{equation}
\label{eq:Slp}
    \mathcal{S}_l(p) = \frac{\mathcal{Q}_l^*(p)}{\mathcal{Q}_l(p)} = \frac{\mathcal{Q}_l(-p)}{\mathcal{Q}_l(p)},
\end{equation}
where the second equality can be shown by invoking the symmetry identities of $\varphi_{l,p}(r)$ and $\hat{h}_l^+(pr)$ consequently establishing the Schwarz reflection principle. 

More importantly, the representation of $\mathcal{S}_l(p)$ in Eq.~\eqref{eq:Slp} reveals that the analytic structure of scattering amplitudes is already encoded at the level of the Schrödinger equation. In particular, poles arising from the zeros of $\mathcal{Q}_l(p)$ persist under analytic continuation and admit a direct physical interpretation as quantum states. This relation identifies the zeros of the Jost function as the source of the $\mathcal{S}$-matrix pole singularities.

\begin{figure*}[!t]
\centering
\includegraphics[width=0.75\linewidth]{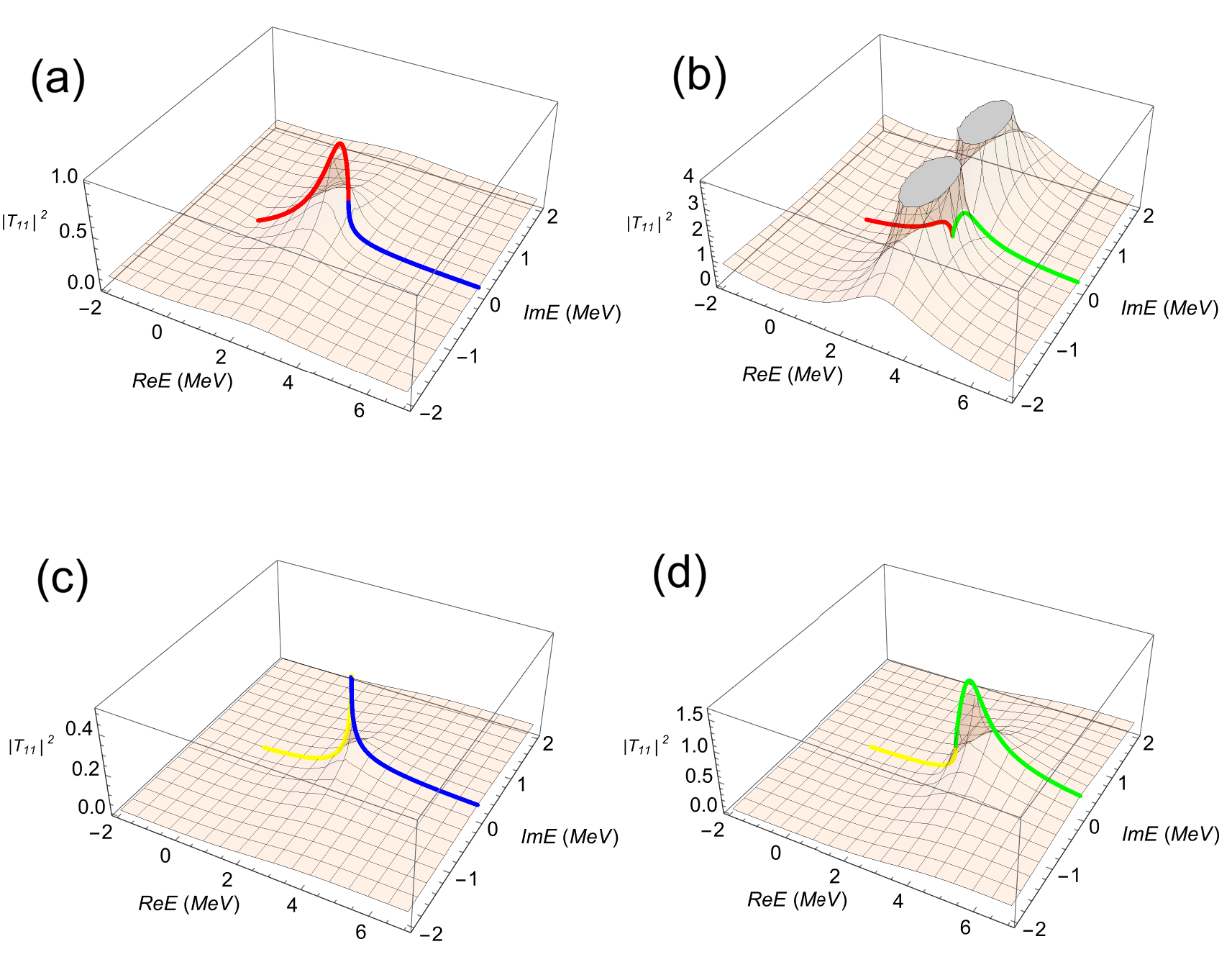}
\caption{Pole structure of a virtual state in the higher-mass channel after strong channel coupling. (a) The physical $[tt]$
 sheet, where the near-threshold enhancement is observed; (b) the $[bt]$ sheet; (c) the $[bb]$ sheet; and (d) the $[tb]$
sheet. The colored lines along the real energy axis indicate the connections among the four Riemann sheets.}
\label{fig:virtualbt}
\end{figure*}

In addition to pole singularities, the branch point associated with the opening of a new scattering channel can also contribute to the complication of interpreting lineshapes. Specifically, if the higher mass channel has a bound state, it can couple to the scattering state of the open channel. In order to illustrate this, we can use the Lippmann-Schwinger equation with separable interaction $v_{m,n}(p,p')=f_{m}(p)\lambda_{mn}f_{n}(p')$ where $\lambda_{mn}$ is the interaction strength coupling of the different channels and $f_{n}(p)$ is the form factor. Here, we follow the treatment done in ~\cite{PearceGibson} where the Yamaguchi form factor for $\ell=0$ is used: $f_{m}(p)=\Lambda_{m}^2/(\Lambda_{m}^2+p^2)$. A virtual state pole of the higher mass channel can be produced by setting $|\lambda_{22}|<2/(\pi\mu_2\Lambda_2)$. Upon turning on the coupled-channel coupling $\lambda_{12}$, this virtual state pole will appear in the $[tb]$ sheet~\cite{Badalyan:1982}. If the coupling is made strong enough, the $[tb]$ sheet pole will
cross the scattering region above the second threshold and end up in the $[bt]$ sheet. This scenario will produce an isolated pole in the $[bt]$ sheet as shown in Fig.~\ref{fig:virtualbt}, which can also be produced by coupling the bound state of the higher mass channel to the lower channel. This result demonstrates the inherent ambiguity in interpreting the observed near-threshold signal due to coupled-channel effects and the influence of the branch point singularity at the threshold.   

Using an interaction model allows us to trace the origin of the poles in the zero-coupling limit. However, it necessarily restricts the interpretation that can be attributed to a near-threshold enhancement. In order to remove this interpretation bias, it is more instructive to use the general features of the $\mathcal{S}$-matrix in generating the lineshape training dataset. 

In a coupled-two-channel system with open channels $\Sigma \,\pi$ and $\bar{K}N$, the $\mathcal{S}$-matrix is a $2\times2$ matrix with elements
\begin{align}
    \mathcal{S}_{11}(p_1, p_2) = \frac{\mathcal{Q}(-p_1, p_2)}{\mathcal{Q}(p_1, p_2)} \quad &, \quad \mathcal{S}_{22}(p_1, p_2) = \frac{\mathcal{Q}(p_1, -p_2)}{\mathcal{Q}(p_1, p_2)} \nonumber \\
    \det \mathcal{S} &= \frac{\mathcal{Q}(-p_1, -p_2)}{\mathcal{Q}(p_1, p_2)}
\end{align}
expressed in terms of the channel momenta;
\begin{equation}
    p_k^2 = \frac{(s-\varepsilon_k^2)[s-\varepsilon_k(\varepsilon_k - 4\mu_k)]}{4s} \simeq s- \varepsilon_k^2,
\end{equation}
where $\varepsilon_k \equiv m_{k1} + m_{k2}$ is the $k^{\mathrm{th}}$ energy threshold, $\mu_k$ is the reduced mass in channel $k$, and $s$ is the Mandelstam variable. For the rest of this manuscript, we utilize the rescaled form $p_k^2 \simeq s - \varepsilon_k^2$ which is motivated by the fact that the analysis focuses on the behavior of the scattering amplitude near the threshold where the $s$-wave contribution dominates and the effects of the higher partial-waves are kinematically suppressed. The presence of kinematical branch points at $\sqrt{s} = \varepsilon_k$ defines the two-channel amplitude on a four-sheeted complex $\sqrt{s}$-plane~\cite{ParticleDataGroup:2024cfk}, which we label as $[tt]$, $[bt]$, $[bb]$, and $[tb]$ sheets following the Pearce and Gibson notation~\cite{PearceGibson}. To systematically analyze the analytic structure without the need to deal with the multivaluedness of the $\mathcal{S}$-matrix in the complex $\sqrt{s}$ variable, we adopt a uniformization scheme following Kato~\cite{Kato:1965iee} that allows the mapping of the four Riemann sheets of the complex $\sqrt{s}$-plane to a single-sheeted cut-free complex $z$-plane via the transformation
\begin{equation}
    z = \frac{p_1 + p_2}{\Delta} \quad ; \quad \frac{1}{z} = \frac{p_1 - p_2}{\Delta},
\end{equation}
where $\Delta \equiv \sqrt{\varepsilon_2^2 - \varepsilon_1^2}$. In this variable, the two-channel $\mathcal{S}$-matrix elements  can be compactly written as
\begin{align}
    \mathcal{S}_{11}(z) = \frac{\mathcal{Q}(-1/z)}{\mathcal{Q}(z)} \quad &, \quad \mathcal{S}_{22}(z) = \frac{\mathcal{Q}(1/z)}{\mathcal{Q}(z)}, \nonumber \\
    \det \mathcal{S} &= \frac{\mathcal{Q}(-z)}{\mathcal{Q}(z)},
\end{align}
where the Jost-like function $\mathcal{Q}(z)$ governs the pole structure of the amplitude. Following~\cite{Kato:1965iee}, we parameterize $\mathcal{Q}(z)$ with simple zeros in complex $z$-plane:
\begin{equation}
\label{eq:Q(z)}
    \mathcal{Q}(z) = \prod^n \frac{1}{z^{2}} (z - z_{n}) (z + z_{n}^{*}) 
    (z - z_{n}') (z + z_{n}^{'*}).
\end{equation}
Here, $n$ denotes the number of physical poles $z_n$ included in the model and $z_n'$ is the regulator pole that ensures the correct asymptotic behavior of the $\mathcal{S}$-matrix elements at high energies. Finally, Schwarz reflection principle requires the introduction of symmetric pole at $-z_n^{(')*}$.

\begin{figure}[!ht]
    \centering
    \includegraphics[width=0.99\linewidth]{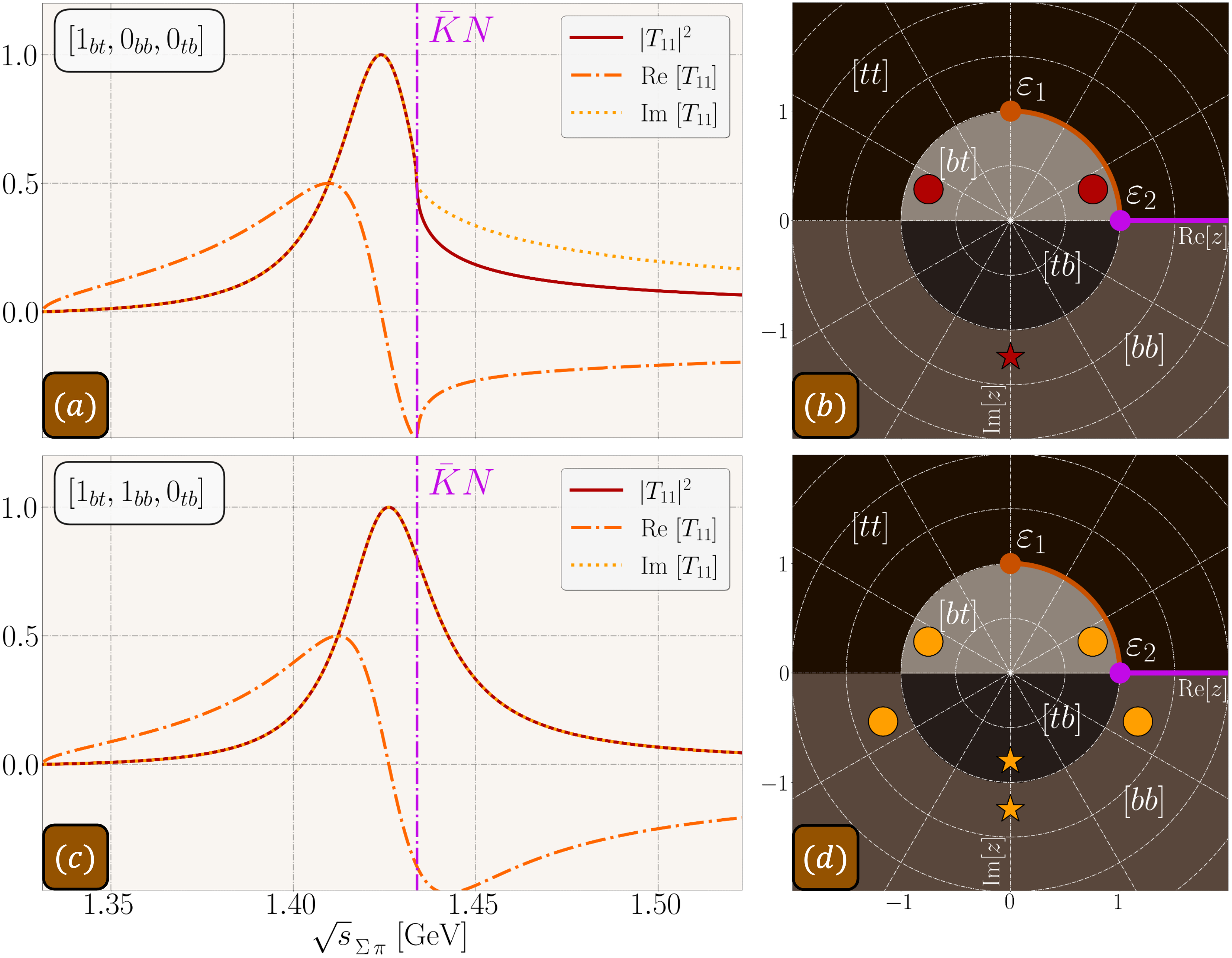}
    \caption{\textbf{Lineshapes of distinct pole structures in cut-free complex $z$-plane.} (a) Scattering lineshapes of an isolated pole on the $[bt]$ sheet and (b) the associated map in the complex $z$-plane. (c) Lineshapes of a pole-shadow configuration and (d) the map of the $z$-plane displaying singularity distribution across the $[bt]$ and $[bb]$ sheets.}
    \label{fig:uni}
\end{figure}

The distribution of these uniformized poles in the complex $z$-plane provide a rigorous diagnostic to discern the underlying physical nature of the hadronic state. By applying the Morgan pole-counting criterion~\cite{Morgan:1992} to the region surrounding the $\bar{K}N$ threshold ($\varepsilon_2$), we can systematically differentiate between a non-molecular state and a dynamically generated hadronic molecule, as illustrated schematically in Figure~\ref{fig:uni}.

When the resonance structure is dynamically generated by the hadron-hadron continuum interaction, the amplitude is typically characterized by a single isolated pole residing on the unphysical $[bt]$ sheet close to $\varepsilon_2$. In the invariant mass spectra, this pole structure yields a highly asymmetric lineshape marked by a sharp cusp at $\sqrt{s} = \varepsilon_2$. This sudden geometric transition is an explicit kinematic threshold effect which is mathematically characterized by a near-infinite slope at the boundary where the new channel opens.

Conversely, an intrinsic compact state introduces an explicit bare degree of freedom. Upon unitarization and coupling to the channels, the state splits into a \textit{pole-shadow pair} distributed symmetrically across both the $[bt]$ and $[bb]$ sheets or adjacent Riemann sheets. This results into a spectral lineshape that is smooth and closely mimics the classic Breit-Wigner distribution.

Crucially, this stark contrast in the localized slopes between different pole structure scenarios provides robust and scale-invariant geometric signature. In this work, this differential feature is utilized as a foundational input for a Convolutional Neural Network (CNN) framework. By training the classifiers on these differential profiles, the models are expected to effectively distinguish bare states from dynamically generated molecular ones, despite the presence of experimental uncertainties and statistical noise. The systematic differential feature formulation and its explicit implementation on the CLAS data is discussed in detail in Section~\ref{sec:3}.


\section{Algorithmic Framework}\label{sec:3}

To resolve the $\Lambda(1405)$ pole structure from scattering data, we move toward a robust machine learning-based methodology. This section details the generation of synthetic dataset, the design of CNN models, and the protocol towards CLAS data inference. 

\subsection{Simulated Data Encoding \label{subsec:traindata}}

Supervised training of deep learning models requires a highly diverse, kinematically consistent, and unambiguously labeled dataset. To satisfy this requirement, we construct synthetic libraries of scattering lineshapes via the uniformized $\mathcal{S}$-matrix discussed in Section~\ref{sec:2}. Because the pole singularities enter the formulation strictly through the zeros of the uniformized Jost-like function $\mathcal{Q}(z)$ in Eq.~\eqref{eq:Q(z)}, each synthesized amplitude can be explicitly mapped to a definitive pole structure label.

In generating the training dataset, we uphold the independence of each assigned pole as described in~\cite{Santos:2023gfh}. Specifically, only the $z_n$ in Eq.~\eqref{eq:Q(z)} is counted as a physical pole while the regulator is set as $z_n'=\exp(-i\pi/2)/|z_n|$. In this construction, the regulator pole has negligible effects in the scattering region, ensuring that only $z_n$ contributes to the relevant structures. This approach removes any interpretative bias that might arise from constraining poles to one another. Consequently, the training dataset consists of lineshapes generated by independent set of poles.

We define $20$ distinct pole structure classes, each characterized by a triplet $[\ell_{bt}^{(1)}, \ell_{bb}^{(2)}, \ell_{tb}^{(3)}]$, where $\ell_{j}^{(i)} \in \{0,1,2,3\}$ denotes the number of poles residing on the $j^{\mathrm{th}}$ Riemann sheet. Notably, poles on the $[tt]$ sheet are excluded \textit{a priori }to satisfy the causality requirement that physical scattering amplitudes are boundary values of functions analytic in the upper-half energy plane~\cite{Screaton:1969he, Minerbo:1971gg} . 

Synthetic amplitudes are evaluated over a discrete energy grid spanning the invariant mass window $[1.330, 1.500]$ GeV. This localized window is chosen to match the critical kinematic region probed by the CLAS Collaboration in their photoproduction measurements of the $\Sigma^+ \pi^-$, $\Sigma^0 \pi^0$, and $\Sigma^- \pi^+$ final states~\cite{CLAS:2013rjt}. To comprehensively account for the energy dependence of the photoproduction mechanism, the analysis incorporates variations across different incoming photon energy intervals within $1.95 < W< 2.85$ [GeV], totaling $27$ distinct empirical datasets across three individual charge channels. While the original measurement extends up to $1.585$ GeV, we truncate our analysis interval at $1.500$ GeV. This intentional restriction focuses the network's attention on the subtle lineshape variations near the $\Sigma\, \pi$ and $\bar{K}N$ thresholds while cleanly eliminating kinematic mixing from the narrow, higher-lying $p$-wave dominated $\Lambda(1520)$ resonance. 

The observable lineshapes are constructed from the differential event rate
\begin{equation}
\label{eq:dN/ds}
    \frac{\mathrm{d}N}{\mathrm{d}\sqrt{s}} = \mathcal{A}\,\left(\left| \sum_{k=1}^{2} \alpha_k \,T_{1k}(\sqrt{s}) \right|^2 + \mathrm{bg}(\sqrt{s})\right),
\end{equation}
where $\mathcal{A}$ is a global normalization constant, $\alpha_k$ are the production coefficients in the elastic and inelastic channels, and $\mathrm{bg}(\sqrt{s})$ is a non-resonant background contribution. The background is randomly selected from two predefined parametrization: a linear function or a vanishing (zero) contribution. In lineshape generation, the pole positions are sampled uniformly within (in GeV)
\begin{equation}
\label{eq:sampspace}
    \begin{cases}
        \,\varepsilon_1 \,\leq \, \text{Re}\, [\sqrt{s}_{\text{pole}}] \, \, \leq 1.5,\\
        0.0 \leq |\text{Im}\, [\sqrt{s}_{\text{pole}}]| \leq 0.5.
    \end{cases}
\end{equation}
This raw synthetic data consists of $i=34$ discrete energy bins $\sqrt{s}_i$ and corresponding intensities $I_i$. While a standard network might look only at the absolute values of $I_i$, the resonance structure of the $\Lambda(1405)$ is uniquely encoded in its differential shape especially near the $\bar{K}N$ threshold. To exploit this, we transform the input into a dual-channel representation that captures the local differential structure of the observable lineshape. 

To provide the network with immediate access to the spectral slope of the lineshape, we compute the discrete finite-difference gradient across adjacent energy points:
\begin{equation}
    \nabla I_i = \frac{I_i - I_{i-1}}{\sqrt{s}_i - \sqrt{s}_{i-1}}, \quad \forall \, i>0.
\end{equation}
To maintain dimensionality and physical continuity across the vector, the slope at the boundary $(i=0)$ is assigned via a forward difference using the first interval so that the resulting vector remains of length $i$. This transformation ensures that the network is sensitive to pole structure features (such as the sharpness of threshold cusps) rather than just the global normalization. 

To mitigate numerical instability arising from near-infinite slopes at the $\bar{K}N$ threshold (for some specific pole topologies), we apply a normalization scheme that maps the differential features into a strictly bounded domain, ensuring that threshold cusps are treated as high-priority signatures rather than numerical singularities. This is achieved via an independent unit-max scaling
\begin{equation}
    \hat{\nabla} I_i = \frac{\nabla I_i }{\mathrm{max}(|\nabla I|)},
\end{equation}
which normalizes the most extreme spectral shift to either $\pm1.0$ (the extrema) and the rest of the slopes to be relative fractions of the extrema. This effectively squashes the cusp into a manageable range that the CNN's filters can process without saturating the activation function. For consistency, the same method was utilized to obtain the normalized intensity values:
\begin{equation}
    \hat{I}_i = \frac{I_i }{\mathrm{max}(I)}.
\end{equation}

Finally, instead of a $1D$ vector, we construct the input tensor $\hat{X}$ by stacking the normalized intensity and finite-difference gradient into a $2\times i$ matrix
\begin{equation}
    \hat{X} = \begin{pmatrix}
        \hat{I}_1 & \hat{I}_2 & \cdots & \hat{I}_i \\
        \hat{\nabla}I_1 & \hat{\nabla}I_2 & \cdots & \hat{\nabla}I_i
    \end{pmatrix},
\end{equation}
which allows the convolutional filters to perform cross-channel correlations. 

Notably, while CNNs are conventionally applied in $2D$ image processing and computer vision, their foundational mechanism (local feature extraction through translation-invariant, weight-sharing kernels) extends naturally to discretized scientific grid data. Interpreting the distinct intensity and slope rows as dual feature maps over a regular energy coordinate system allows the CNNs to adapt their spatial pattern-recognition capabilities to physical scattering lineshapes.

By sliding a kernel over both rows simultaneously, the models learn to identify \textit{slope-intensity} signatures, such as patterns where a high intensity is coupled with a sharp negative slope, indicating the presence of a near-threshold pole.

\subsection{Ensemble Architecture Design}
To ensure that the identification of the $\Lambda(1405)$ pole structure is invariant to specific model initializations or architectural biases, we implement a heterogeneous ensemble of  deep convolutional neural networks. This ensemble approach allows for a probabilistic interpretation of the classification task, where the final prediction is derived from a consensus of diverse learners. To achieve a robust ensemble, we systematically build a family of constituent models designed to process and interpret the input space through varying the hyperparameters via the Optuna optimization framework~\cite{optuna_2019}. The hyperparameter search space explored during optimization spans both the convolutional base and the fully connected dense layers, as outlined in Table~\ref{tab:hyp_space}.
\begin{table}[!ht]
\centering
\setlength{\tabcolsep}{8pt}
\renewcommand{\arraystretch}{1.5}
\caption{Hyperparameter search space explored by Optuna optimizer~\cite{optuna_2019} for the constituent ensemble architectures.}
\label{tab:hyp_space}
\begin{tabular}{llc}
\hline\hline
\textbf{Hyperparameter} & \textbf{Symbol} & \textbf{Domain} \\
\hline
Kernel size 1 & $k_1$\textsuperscript{*} & $\{3, 5, 7\}$ \\
Dilation rate 1 & $d_1$\textsuperscript{*} & $\{1, 2\}$ \\
Kernel size 2 & $k_2$\textsuperscript{*} & $\{3, 5, 7\}$ \\
Dilation rate 2 & $d_2$\textsuperscript{*} & $\{1, 2\}$ \\
Spatial dropout & $p_{\text{s}}$\textsuperscript{**} & $[0.0, 0.3]$ \\
FC depth & $h_1$\textsuperscript{***} & $[2, 4]$ \\
FC units & $h_2$\textsuperscript{***} & $\{64x\}_{x=1}^8
$ \\
FC dropout rate & $p_i$\textsuperscript{**} & $[0.05, 0.40]$ \\
Focal gamma & $\gamma_{\,\text{F.L.}}$\textsuperscript{**} & $[1.0, 3.0]$ \\
Max. learning rate & $\text{LR}_{\text{max}}$\textsuperscript{**} & $[10^{-5}, 10^{-3}]$ \\
Weight decay & $\omega_d$\textsuperscript{**} & $[10^{-6}, 10^{-2}]$ \\
\hline\hline
\multicolumn{3}{l}{\footnotesize \textsuperscript{*}Categorical \quad \textsuperscript{**}Continuous float \quad \textsuperscript{***}Discrete integer}
\end{tabular}
\end{table}

The search space was explored using Optuna's structured Parzen estimator algorithm paired with an automated \texttt{MedianPruner}. In the screening phase, trial architectures were evaluated based on their accuracy after epoch $50$, with underperforming trials pruned. To prevent early termination of slow-starting CNN models, a warmup period of $15$ epochs was enforced before evaluating trials against the median benchmark.

By varying the kernel sizes alongside explicit dilation rates~\cite{Yu:2016}, the architecture shifts how it samples the $34$ energy bins. Models with small kernels are restricted to narrow spatial fields which forces them to map localized geometric variations and sharp threshold cusps with high fidelity. In contrast, models incorporating larger kernels or increased dilation rates expand their field of view without introducing additional parameters which allows them to capture long-range spectral correlations and the global lineshape of the resonance. For the convolutional layers to look at both the intensity and slope arrays independently, spatial dropout~\cite{Tompson:2015} was introduced to drop some feature maps. 

Beyond the convolutional base, the architectural variance is maintained by systematically scaling the depth and width of the fully-connected (FC) hidden layers paired with shifting regularization profiles. Dense layers with lower capacity restrict the network to learning only the most dominant physical patterns, whereas wider and deeper architectures provide the necessary capacity to separate highly degenerate lineshapes. This capacity scaling is counterbalanced by varying the dropout rates~\cite{Srivastava:2014}. Higher dropout values prevent co-adaptation by forcing the dense layers to discover redundant, independent pathways to a classification, while lower dropout values allow the models to remain highly sensitive to faint physical patterns.

The entire ensemble framework is implemented using the PyTorch package~\cite{Paszke:2019xhz}. To avoid the dying ReLU problem during backpropagation, every architecture utilizes the LeakyReLU activation functions~\cite{Maas:2013, Xu:2015}. Moreover, to prioritize near-degenerate lineshapes, the models are trained using a multi-class Focal Loss (F.L.) objective function~\cite{Lin:2017}. The focal parameter $\gamma_{\,\mathrm{F.L.}}$ is customized for each network to control the down-weighting of easily classified lineshape samples and force each CNN to focus their learning capacity on highly distorted physical profiles.

The optimization process is driven by the AdamW optimizer~\cite{Loshchilov:2019}, which decouples weight decay from the primary gradient updates to ensure stable generalization across the architectures. Finally, to accelerate convergence and avoid local minima, the learning rate for each model is  regulated via a OneCycleLR scheduler~\cite{Smith:2019}. This dynamically cycles the optimization path up to a designated maximum learning rate (Max LR) before performing a cosine annealing down to a minimum threshold which regularizes the training procedure and smooths the final consensus boundary.

\subsection{CLAS Data Inference Protocol}

In the inference phase, the CNN ensemble was utilized to evaluate the pole structure of the $\Lambda(1405)$ across $27$ empirical datasets from CLAS Collaboration~\cite{CLAS:2013rjt}. This comprises $9$ distinct kinematic bins evaluated across three charge channels: $\Sigma^+\pi^-$, $\Sigma^0 \pi^0$, and $\Sigma^- \pi^+$. In Sec.~\ref{sec:4}, the datasets are analyzed both individually through separate charge channel analyses and collectively through a unified analysis, in which the lower channel is treated as the isospin-averaged $\Sigma\, \pi$ system. To rigorously account for the experimental uncertainties, each empirical dataset is subjected to Monte Carlo noise-perturbation framework. For a given invariant mass datum with reported experimental error, we generate $10^4$ synthetic variations by sampling the error band via a Gaussian distribution. For every generated perturbation, a dual-array feature map is constructed by stacking the normalized intensity array alongside its first-order spatial numerical derivative. To be consistent with our training dataset generation protocol (as described in Sec.~\ref{subsec:traindata}), both the intensity and slope arrays are independently mapped to a $[-1, 1]$ bound by dividing by their respective absolute maximum values.

The inputs are processed by all ensemble members concurrently. With the goal to eliminate the risk of systematic architectural bias and decouple the final statistical analysis from runtime deep-learning frameworks, each model maps its categorical logits directly to a discrete voting tally vector corresponding to the $20$ unique pole configurations. The global ensemble consensus is quantified by evaluating the mean vote support and the epistemic uncertainty of the consensus is derived from the ensemble variance of the mean.


\section{\texorpdfstring{$\Lambda(1405)$}{} Pole Structure \label{sec:4}}

To investigate the robustness of the inferred pole structure, the CLAS data are analyzed using two complementary approaches. The first is an independent analysis, in which each charge channel ($\Sigma^+\pi^-$, $\Sigma^0 \pi^0$, and $\Sigma^- \pi^+$) is analyzed separately using their corresponding physical thresholds. The second is a unified analysis, in which the three $\Sigma\, \pi$ charge channels are treated collectively by adopting an isospin-averaged lower threshold. 

\subsection{Independent analysis of \texorpdfstring{$\Sigma \, \pi$}{} charge channels}
\begin{figure*}[!ht]
    \centering
    \includegraphics[width=0.98\linewidth]{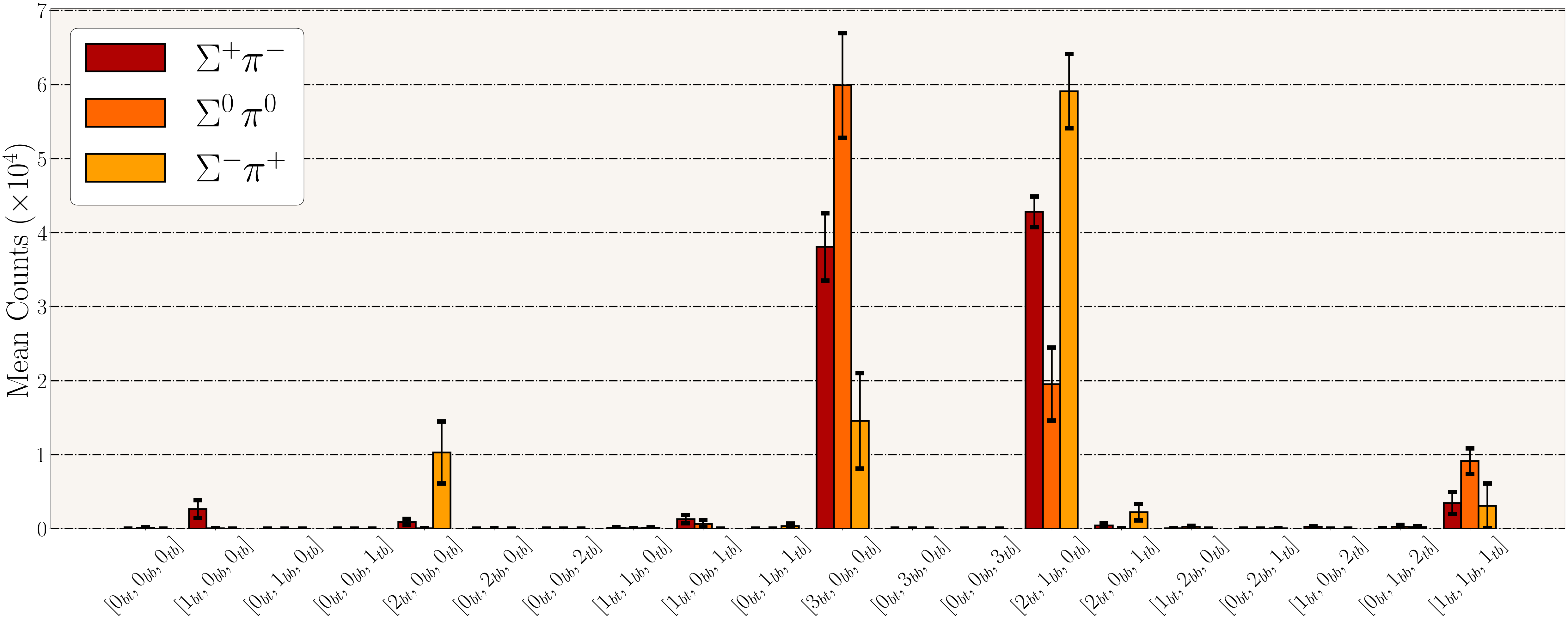}
    \caption{\textbf{Predicted class distribution from the independent analysis.} Deep learning ensemble consensus on the $\Lambda(1405)$ pole structure. Classification distributions are generated using $3$ independently initialized CNN classifiers evaluated on the CLAS data. The error bars represent the standard error of the mean.}
    \label{fig:PCD_IA}
\end{figure*}
To isolate final-state interactions, charge channel-specific lineshape distortions, and kinematic threshold splittings without imposing global averaging, each channel was evaluated independently. For each charge channel, a dedicated $3$-member CNN ensemble was constructed using Optuna~\cite{optuna_2019}. The hyperparameter configurations for these charge channel-specific architectures are detailed in Table~\ref{tab:CNNE_IA}.
\begin{table*}[!ht]
    \centering
    \setlength{\tabcolsep}{7pt}
    \renewcommand{\arraystretch}{1.5}
    \caption{\textbf{Independent analysis.} Hyperparameter configurations for the 3-member CNN architectures via Optuna optimizer~\cite{optuna_2019}.}
    \label{tab:CNNE_PA}
    \begin{tabular}{ccccccccc} 
    \hline \hline
       Model & $[k_1, k_2]$ & $[d_1, d_2]$ & Hidden Layers & $p_i$ & $p_{\mathrm{s}}$ & $\mathrm{LR}_{\mathrm{max}}$ & $\gamma_{\,\mathrm{F.L.}}$ & $\omega_d$\\
    \hline \hline
    $\Sigma^+ \pi^-$ &  &  &  &  &  &  &  &  \\
    $\mathrm{I}$ & $[5,3]$ & $[1,2]$ & $[256, 384]$ & $[0.23, 0.06]$ & $0.048$ & $4.44\times 10^{-4}$ & $1.81$ & $9.15 \times 10^{-4}$ \\
    $\mathrm{II}$ & $[7,5]$ & $[2,1]$ & $[448, 64, 384]$ & $[0.10, 0.25, 0.18]$ & $0.276$ & $8.81 \times 10^{-4}$ & $2.17$ & $2.03 \times 10^{-2}$ \\
    $\mathrm{III}$ & $[5, 3]$ & $[1,1]$ & $[384, 512, 64, 256]$ & $[0.35, 0.33, 0.06, 0.32]$ & $0.295$ & $9.23 \times 10^{-3}$ & $2.78$ & $8.69 \times 10^{-4}$ \\
    \hline
    $\Sigma^0 \pi^0$ &  &  &  &  &  &  &  &  \\
    $\mathrm{I}$ & $[5,5]$ & $[1,1]$ & $[128, 128]$ & $[0.39, 0.31]$ & $0.178$ & $1.29 \times 10^{-3}$ & $1.51$ & $1.26 \times 10^{-3}$ \\
    $\mathrm{II}$ & $[7,5]$ & $[1,1]$ & $[384, 64, 256]$ & $[0.24, 0.21, 0.25]$ & $0.148$ & $4.86 \times 10^{-4}$ & $1.05$ & $2.35 \times 10^{-2}$ \\
    $\mathrm{III}$ & $[7, 7]$ & $[1,2]$ & $[128, 512, 512, 64]$ & $[0.32, 0.32, 0.35, 0.17]$ & $0.185$ & $3.66 \times 10^{-3}$ & $1.77$ & $1.62 \times 10^{-3}$ \\
    \hline
    $\Sigma^- \pi^+$ &  &  &  &  &  &  &  &  \\
    $\mathrm{I}$ &$[3,3]$ & $[1,1]$ & $[448, 192]$ & $[0.18, 0.13]$ & $0.083$ & $2.09 \times 10^{-4}$ & $1.78$ & $2.23 \times 10^{-5}$ \\
    $\mathrm{II}$ & $[3,3]$ & $[2,1]$ & $[192, 256, 512]$ & $[0.11, 0.27, 0.20]$ & $0.231$ & $8.82 \times 10^{-4}$ & $1.31$ & $1.91\times 10^{-4}$ \\
    $\mathrm{III}$ & $[7, 3]$ & $[2,2]$ & $[64, 384, 320, 192]$ & $[0.24, 0.29, 0.28, 0.26]$ & $0.156$ & $3.11 \times 10^{-3}$ & $1.60$ & $1.69 \times 10^{-2}$ \\
    \hline \hline
    \end{tabular}
\end{table*}
As shown, each charge channel-specific ensemble implements a progressive depth hierarchy. All of these models were trained up to $300$ epochs and achieved stable convergence with both training and validation accuracies consistently falling within the range of $85\%$ to $89\%$, implying that each CNN model achieved high expressiveness but still maintain strong generalization without overfitting.

Evaluating the trained CNNs against $10^4$ bootstrapped empirical lineshapes per dataset across each charge channel (which sums up to $9\times10^4$ empirically-derived inference lineshapes) yields the predicted class support distribution shown in Fig.~\ref{fig:PCD_IA}. Most notably, across all three channels, the single pole topologies receive near-zero statistical support. Regardless of differences in localized lower channel thresholds, the models unambiguously reject single-pole topologies in favor of multi-pole configurations. While multi-pole configuration universally dominates, the dominant preference rotates between these charge channels. In $\Sigma^+ \pi^-$ channel, CNN prediction is shared almost equally between the $[2_{bt}, 1_{bb}, 0_{tb}]$ and $[3_{bt}, 0_{bb}, 0_{tb}]$ pole topologies. For the datasets on the $\Sigma^0 \pi^0$ channel, the CNNs' predictions strongly favors the $[3_{bt}, 0_{bb}, 0_{tb}]$ configuration with $[2_{bt}, 1_{bb}, 0_{tb}]$ serving as a secondary candidate. In $\Sigma^- \pi^+$ channel, the $[2_{bt}, 1_{bb}, 0_{tb}]$ configuration was decisively favored while $[3_{bt}, 0_{bb}, 0_{tb}]$ becomes the secondary candidate. 

Clearly, evaluating each charged channels in isolation led to ambiguous classification results due to varying empirical statistics. Building upon these charge channel-specific insights, we now transition to unified global analysis to synthesize these localized dynamics into isospin-averaged consensus.

\subsection{Unified analysis of \texorpdfstring{$\Sigma \, \pi$}{} charge channels}
\begin{figure*}[!ht]
    \centering
    \includegraphics[width=0.98\linewidth]{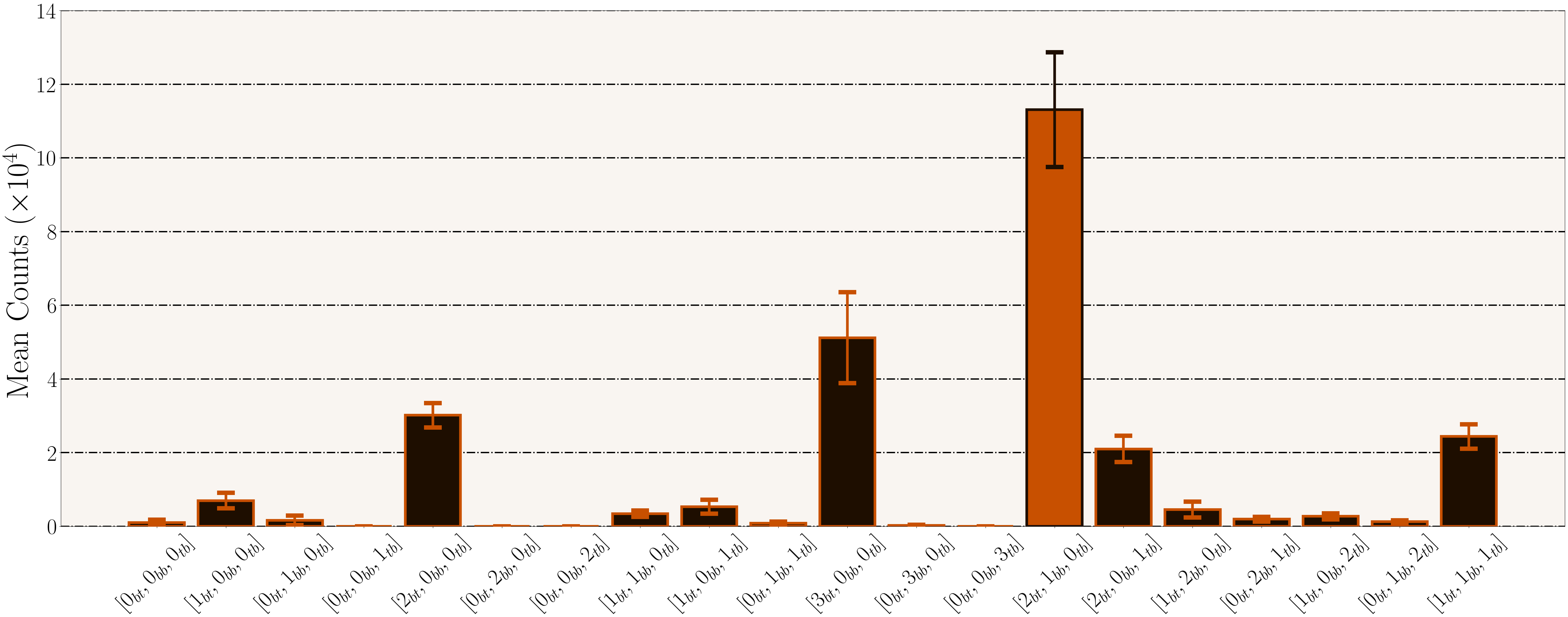}
    \caption{\textbf{Predicted class distribution from the unified analysis.} Deep learning ensemble consensus on the $\Lambda(1405)$ pole structure. Classification distributions are generated using $6$ independently initialized CNN classifiers evaluated on the CLAS data. The error bars represent the standard error of the mean.}
    \label{fig:PCD_UA}
\end{figure*}

\begin{table*}[!ht]
    \centering
    \setlength{\tabcolsep}{7pt}
    \renewcommand{\arraystretch}{1.5}
    \caption{\textbf{Unified analysis.} Hyperparameter configurations for the 6-member CNN architectures via Optuna optimizer~\cite{optuna_2019}.}
    \label{tab:CNNE_IA}
    \begin{tabular}{ccccccccc} 
    \hline \hline
       Model & $[k_1, k_2]$ & $[d_1, d_2]$ & Hidden Layers & $p_i$ & $p_{\mathrm{s}}$ & $\mathrm{LR}_{\mathrm{max}}$ & $\gamma_{\,\mathrm{F.L.}}$ & $\omega_d$\\
    \hline \hline
    $\mathrm{I}$ & $[5,5]$ & $[2,1]$ & $[512, 64, 64, 384]$ & $[0.30, 0.33, 0.08, 0.22]$  & $0.112$ & $2.72\times10^{-3}$ & $2.04$ & $1.39 \times10^{-5}$ \\
    $\mathrm{II}$ & $[7,7]$ & $[2,1]$ & $[64, 64, 448]$ & $[0.18, 0.08, 0.30]$ & $0.015$ & $2.01 \times 10^{-3}$ & $1.45$ & $1.72 \times 10^{-4}$ \\
    $\mathrm{III}$ & $[7,5]$ & $[2,2]$ & $[448, 256, 128]$ & $[0.39, 0.20, 0.33]$ & $0.122$ & $2.94 \times 10^{-3}$ & $2.02$ & $5.25 \times 10^{-4}$ \\
    $\mathrm{IV}$ & $[5,7]$ & $[2,2]$ & $[256, 256]$ & $[0.36, 0.19]$ & $0.166$ & $6.24\times10^{-4}$ & $2.66$ & $1.93\times10^{-5}$ \\
    $\mathrm{V}$ & $[5,3]$ & $[1,2]$ & $[256, 448]$ & $[0.31, 0.31]$ & $0.108$ & $3.30 \times 10^{-3}$ & $1.55$ & $2.86 \times 10^{-4}$ \\
    $\mathrm{VI}$ & $[5,3]$ & $[1,2]$ & $[448, 64, 512, 384]$ & $[0.32, 0.26, 0.28, 0.10]$ & $0.192$ & $4.32\times 10^{-3}$ & $2.81$ & $1.49 \times 10^{-3}$ \\
    \hline \hline
    \end{tabular}
\end{table*}

Following the independent charge-channel analyses, we perform a unified global analysis by treating the lower channel in the isospin basis using an isospin-averaged $\Sigma \,\pi$ threshold. For this unified inference, we employed a $6$-member independently trained CNNs, whose hyperparameters are summarized in Table~\ref{tab:CNNE_IA}. As demonstrated, this $6$-member ensemble spans broad architectural diversity. Models $\mathrm{IV}$ and $\mathrm{V}$ provide lightweight $2$-layer dense configurations; models $\mathrm{II}$ and $\mathrm{III}$ establish intermediate $3$-layer representations; and models $\mathrm{I}$ and $\mathrm{VI}$ deploy deep $4$-layer dense networks scaling up to $512$ hidden units. The convolutional kernels exhibit larger average spatial dimensions paired with non-trivial dilation rates ($d_i =2$) which gives the networks broad receptive field to capture long-range spectral features across the energy domain. The focal loss parameters ($\gamma_{\mathrm{\,F.L.}}$) that ranges from $1.45$ to $2.81$ maintains targeted sensitivity along ambiguous lineshapes during the full $500$-epoch training process.

To perform global inference on CLAS data, $10^4$ empirically-derived lineshapes were generated by normally sampling the experimental uncertainties of each datum. This procedure yields a total evaluation corpus of $27 \times10^{4}$ bootstrapped CLAS lineshapes. Unlike the independent analysis, all empirically-derived lineshapes were simultaneously fed to each trained CNNs to infer the most probable pole structure that best describes the complete CLAS dataset. Consequently, the inferred pole structure represents a global probabilistic interpretation of the $\Lambda(1405)$ analytic structure, rather than one associated with a particular charge channel or photon energy bin.

The global candidate class support distribution inferred by the CNN ensemble is summarized in Fig.~\ref{fig:PCD_UA}. The pole structure class $[2_{bt}, 1_{bb}, 0_{tb}]$ emerges as the indisputable global preference, accumulating $\sim 11.3 \times 10^4$ mean counts (about $40\%$ of the total probability across the $27\times10^4$ inference samples). This configuration asserts that the empirical lineshapes are governed by $2$ dominant poles on the $[bt]$ sheet accompanied by a pole on the $[bb]$ sheet. However, non-negligible support is also assigned to the $[3_{bt}, 0_{bb}, 0_{tb}]$, $[2_{bt}, 0_{bb}, 0_{tb}]$,  $[1_{bt}, 1_{bb}, 1_{tb}]$, and  $[2_{bt}, 0_{bb}, 1_{tb}]$ pole structure classes. Remarkably, with the sole exception of the pole structure $[1_{bt}, 1_{bb}, 1_{tb}]$\footnote{Even this pole structure cannot be simply classified as a traditional single-pole model on the primary $[bt]$ sheet, as strong interchannel coupling allows the pole on the $[tb]$ sheet to dynamically traverse to the $[bt]$ sheet, as demonstrated in various model-dependent analyses~\cite{Hanhart:2014ssa, PearceGibson, Frazer:1964, Badalyan:1982}.}, these configurations that contains at least $2$ poles on the $[bt]$ sheet ($n_{bt} \geq2$) account for the vast majority of the votes. This reveals a central physical result: the presence of $2$ poles on the $[bt]$ sheet is remarkably robust, whereas the precise existence or sheet location of the third pole remains the primary source of residual ambiguity. Nonetheless, conventional and popular single-pole model for the $\Lambda(1405)$ is unambiguously ruled out by the CNN models at this point.

Crucially, however, a definitive physical interpretation cannot be assigned solely on the basis of this global classification. Naively, the dominant $[2_{bt}, 1_{bb}, 0_{tb}]$ pole structure could be interpreted as a two-state system consisting of a single hadronic molecular state ($1$ pole in the $[bt]$ sheet) alongside a genuine resonant state (pole-shadow pair in the $[bt]-[bb]$ sheets). Nonetheless, a standalone pole structure identification does not specify how these poles are distributed in the complex energy plane. Without precise pole locations, one cannot determine whether some of the singularities act as a dynamically generated quantum states or simply compensating for subtle background specific patterns in the empirical data. Disentangling these physical mechanisms and mapping the exact spectrum of $\Lambda(1405)$ directly warrants the quantitative extraction of pole parameters, which we turn in the next section.

\begin{figure*}[!ht]
    \centering
    \subfloat{\includegraphics[width=0.32\linewidth]{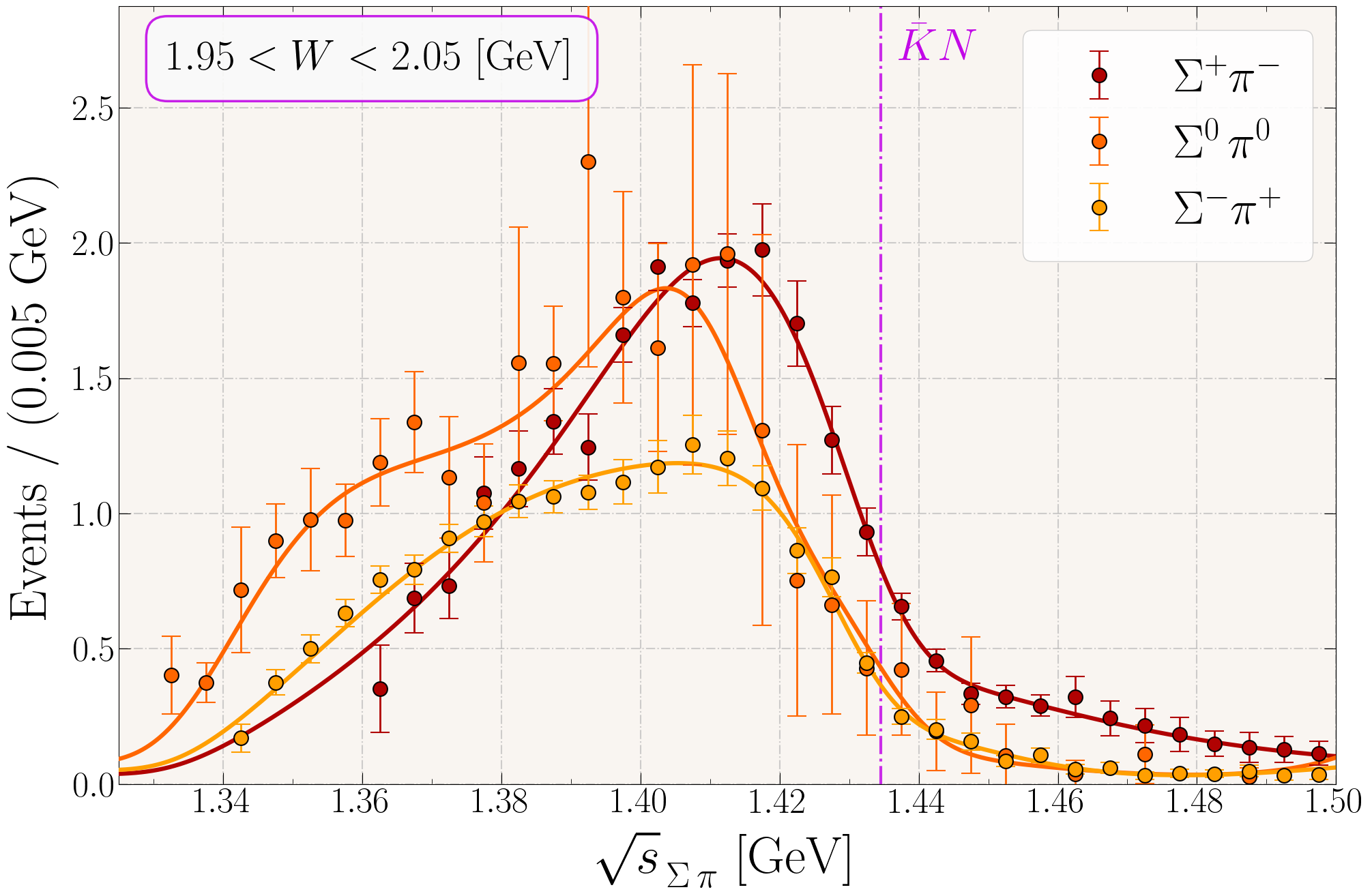}}
    \subfloat{\includegraphics[width=0.32\linewidth]{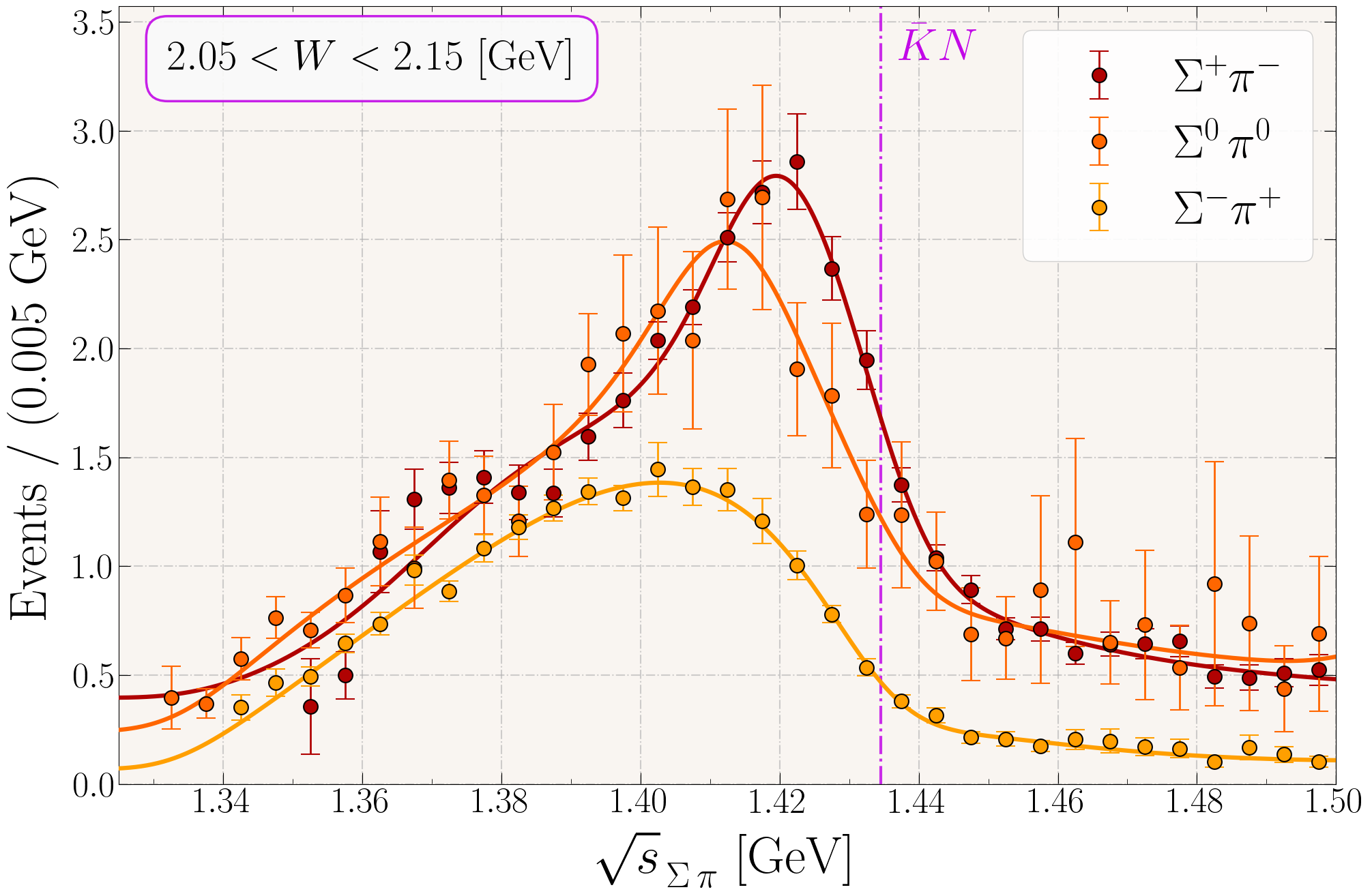}}
    \subfloat{\includegraphics[width=0.32\linewidth]{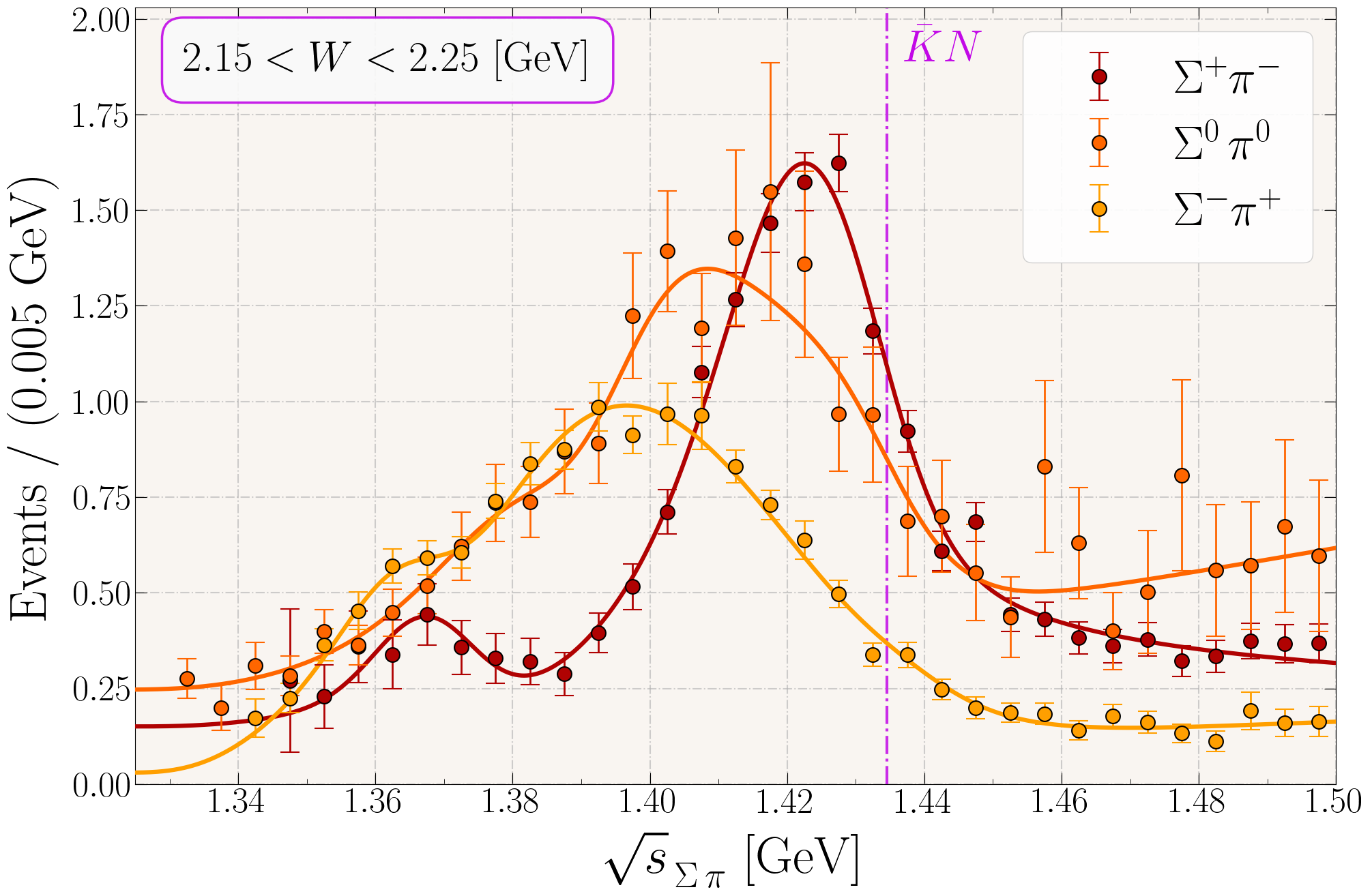}} 
    \\
    \subfloat{\includegraphics[width=0.32\linewidth]{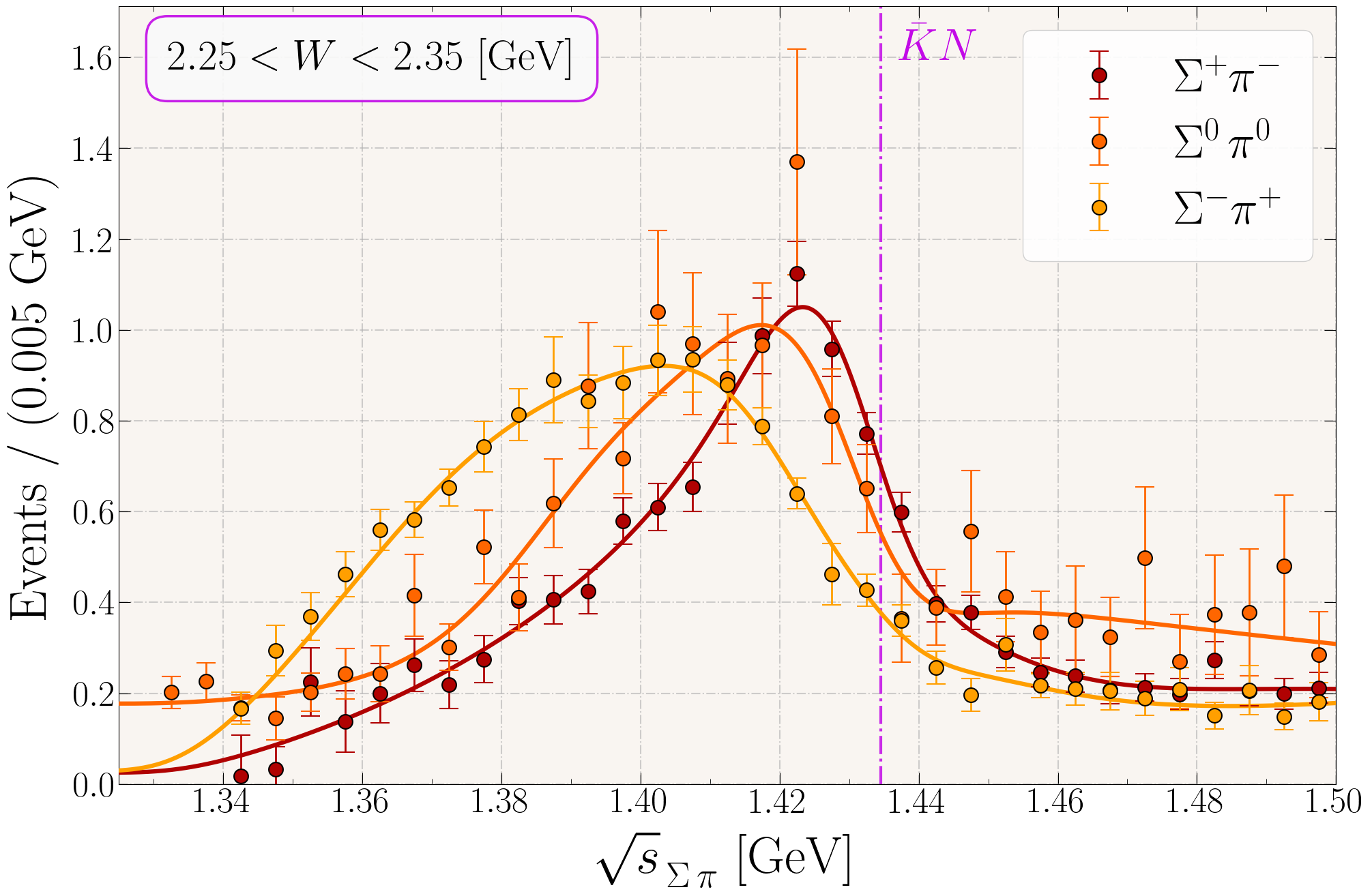}}
    \subfloat{\includegraphics[width=0.32\linewidth]{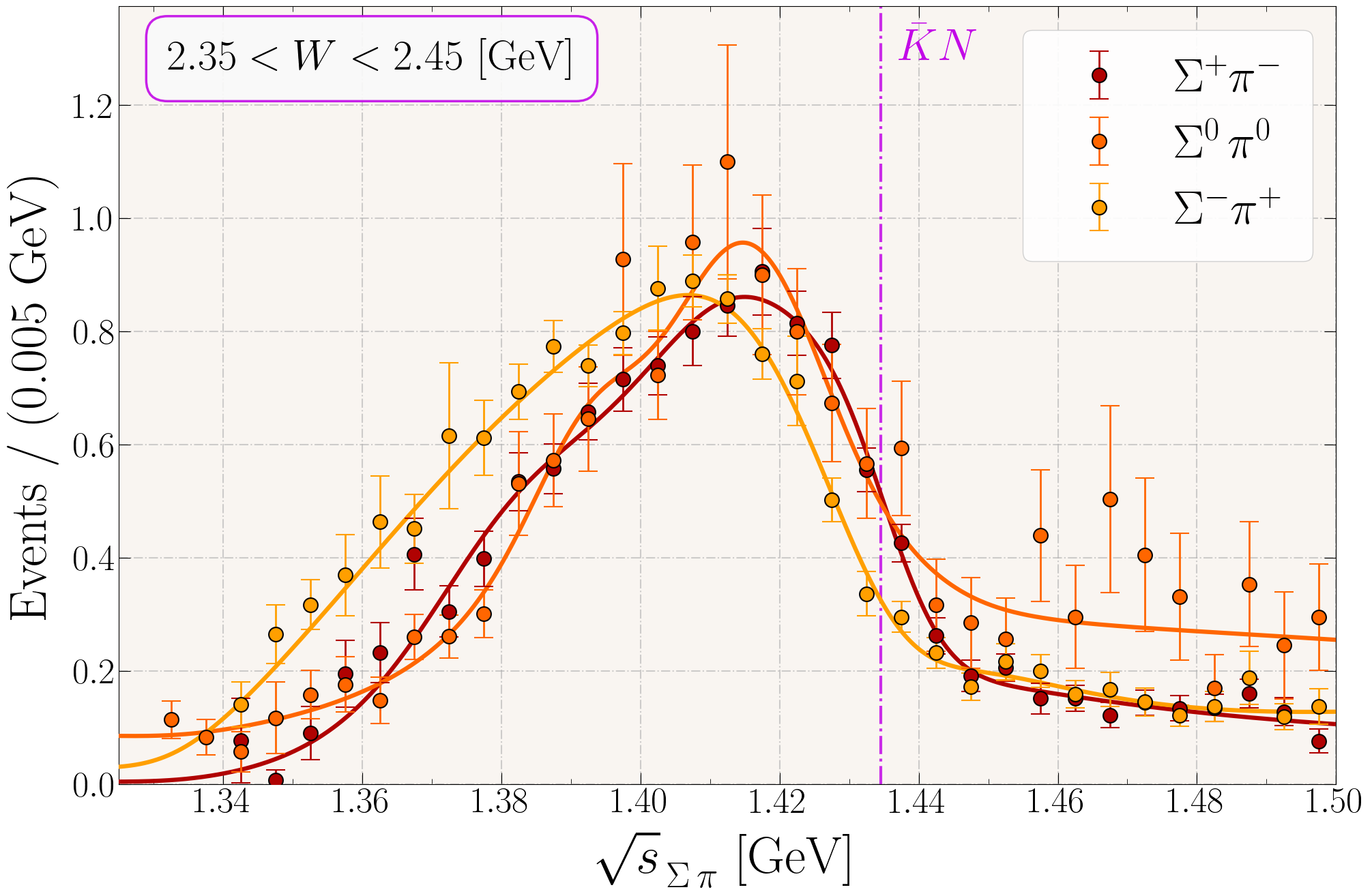}}
    \subfloat{\includegraphics[width=0.32\linewidth]{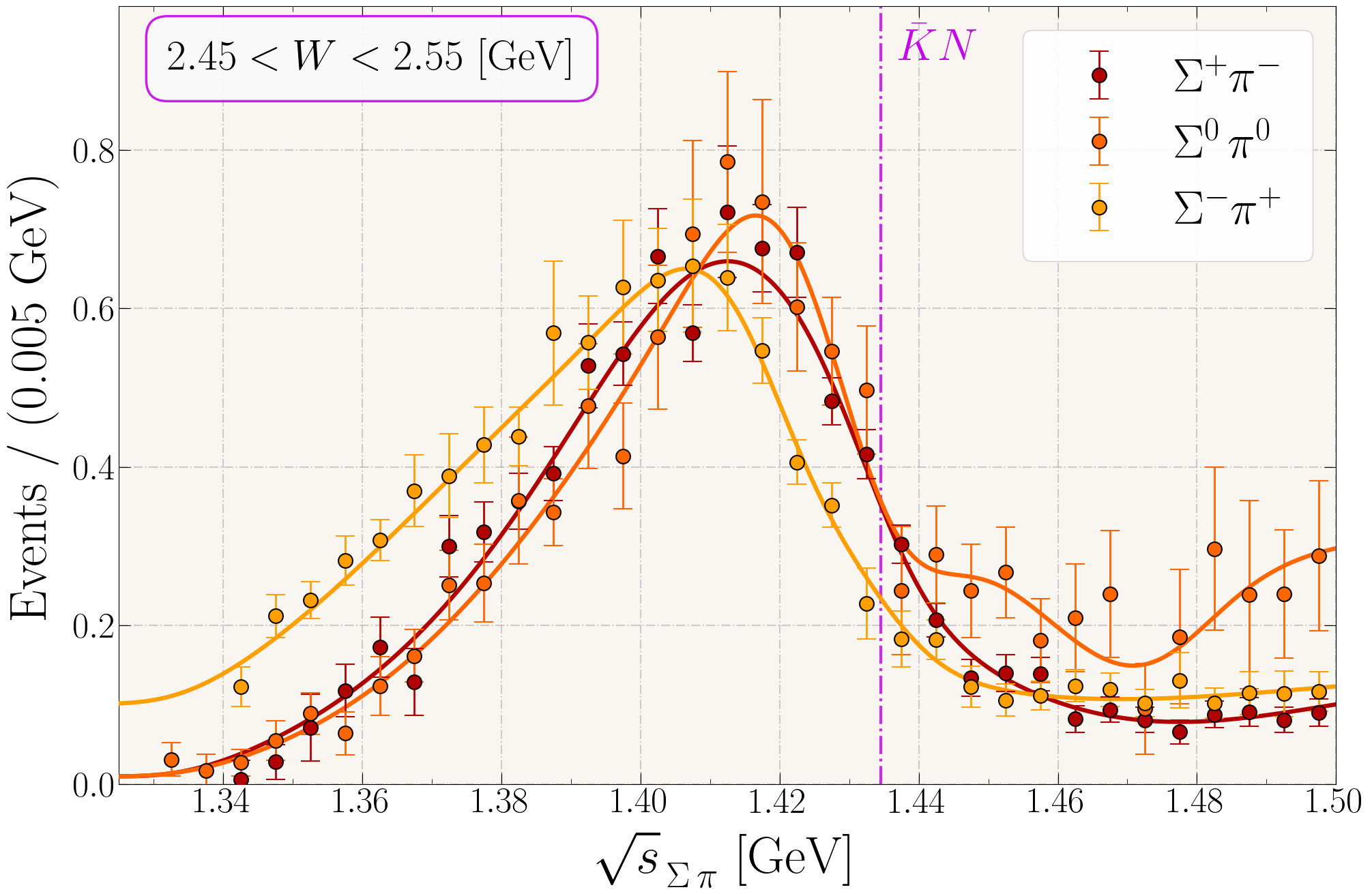}}
    \\
    \subfloat{\includegraphics[width=0.32\linewidth]{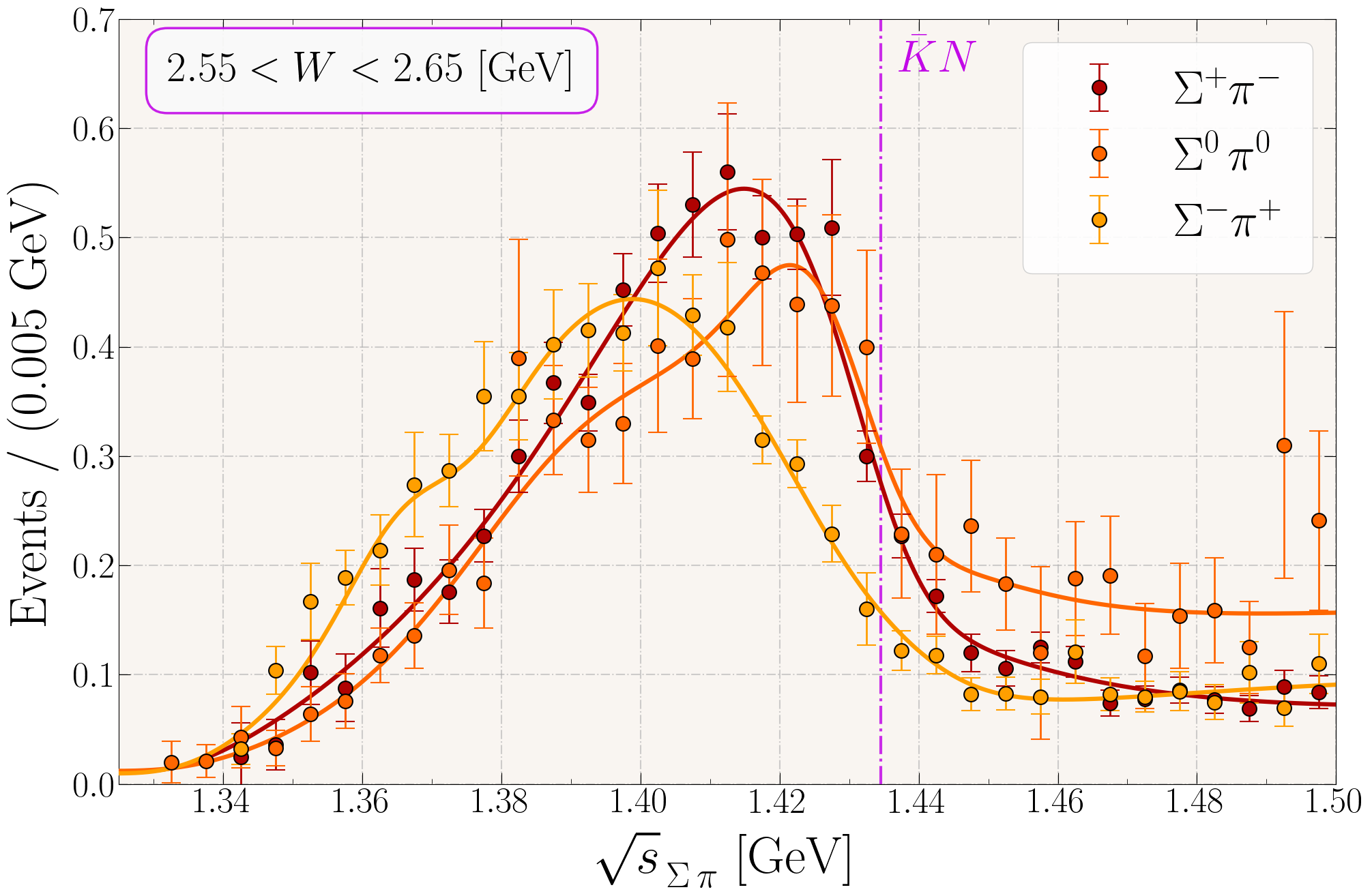}}
    \subfloat{\includegraphics[width=0.32\linewidth]{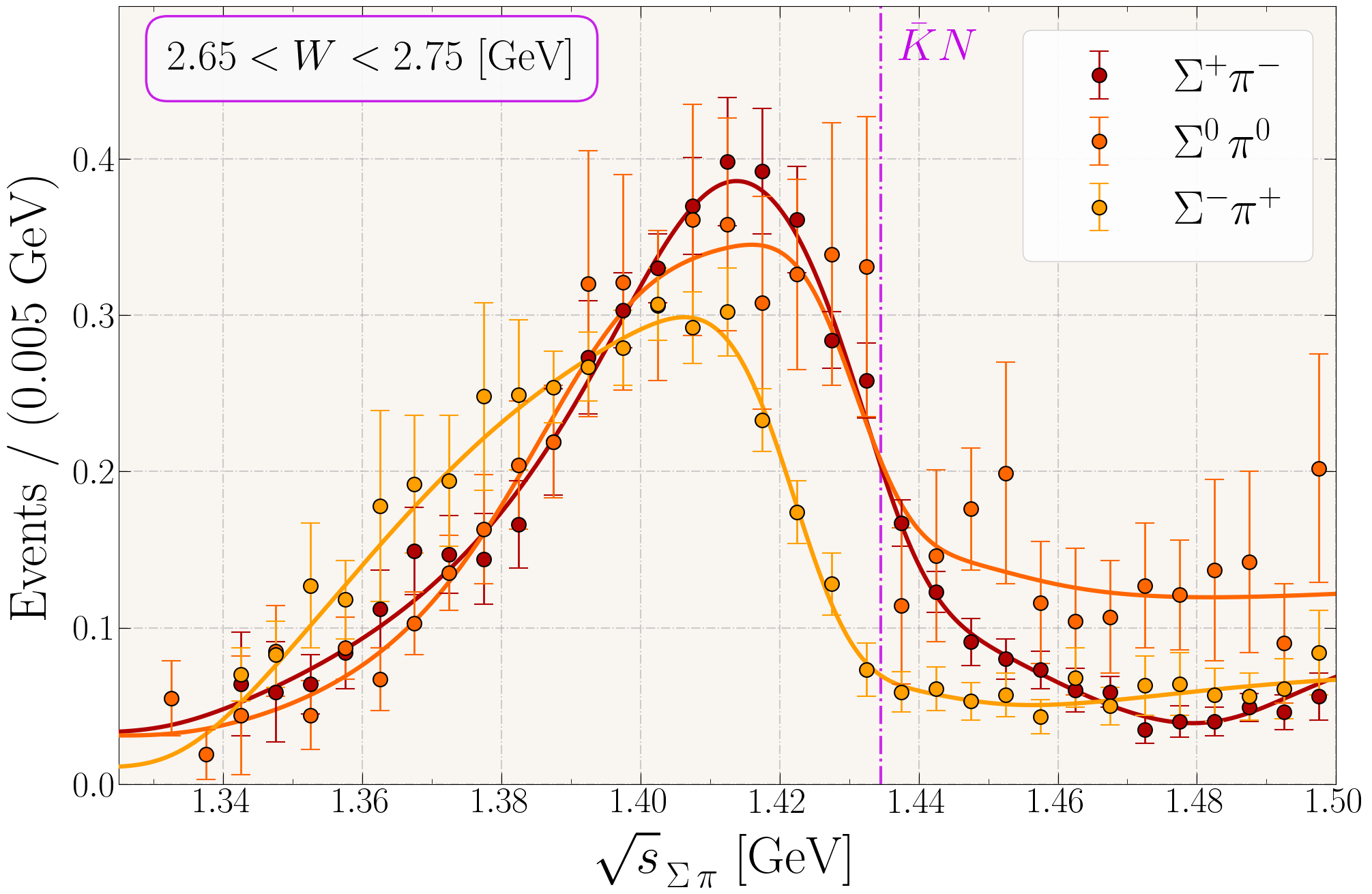}}
    \subfloat{\includegraphics[width=0.32\linewidth]{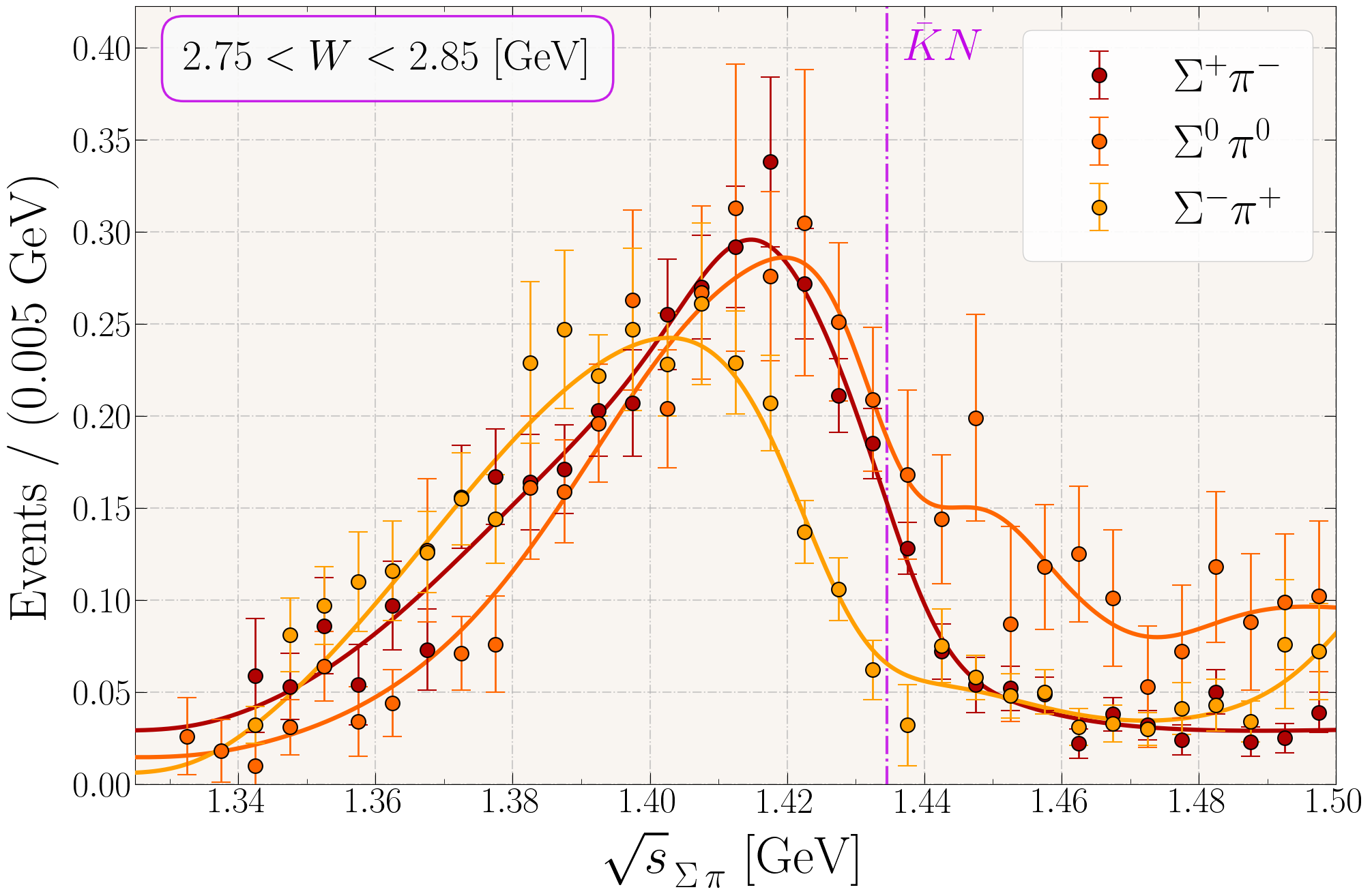}}
    \caption{\textbf{Best fit lineshapes for the $[2_{bt}, 1_{bb}, 0_{tb}]$ model.}}
    \label{fig:minuitfit}
\end{figure*}

\begin{table*}[!ht]
    \centering
    \setlength{\tabcolsep}{15pt}
    \renewcommand{\arraystretch}{2.0}
    \caption{\textbf{Extracted pole parameters for the multi-pole structure of the $\Lambda(1405)$}. Mean values and standard error of the mean for the real ($M$) and imaginary ($\Gamma/2$) parts of the pole position are extracted from multiple COM energy intervals $W$ and decay channels ($\Sigma^+\pi^-$, $\Sigma^0 \, \pi^0$, and $\Sigma^- \pi^+$).}
    \begin{tabular}{ccc}
    \hline \hline
        $\sqrt{s}_{\,[bt],1}^{\mathrm{(pole)}}$ [MeV] & $\sqrt{s}_{\,[bt],2}^{\mathrm{(pole)}}$ [MeV] & $\sqrt{s}_{\,[bb]}^{\mathrm{(pole)}}$ [MeV]\\
        \hline
         $(1422.48 \pm 2.36) - (25.45 \pm 3.83)i$ & $(1380.32 \pm 3.46) - (85.64 \pm 13.32)i$ & $(1386.32 \pm 10.29) - (111.74 \pm 20.79)i$\\
         \hline \hline
    \end{tabular}
    \label{tab:poleparmean}
\end{table*}

\section{Pole Parameters \label{sec:5}}

Guided by the CNN ensemble classification established in the preceding section, we now perform a quantitative extraction of the complex pole positions governing the $\Lambda(1405)$ energy region. The inference-guided fitting procedure directly adopts the dominant $[2_{bt}, 1_{bb}, 0_{tb}]$ pole structure as the baseline analytic ansatz for the empirical CLAS lineshapes. Utilizing the generalized $\mathcal{S}$-matrix parametrization formalism detailed in Sec.~\ref{sec:2}, a pole on the $j^{\mathrm{th}}$ sheet is directly parameterized as
\begin{equation}
    \sqrt{s}_{\,[j]}^{\mathrm{(pole)}} = M - \frac{i\,\Gamma}{2},
\end{equation}
where $M$ corresponds to the pole mass and $\Gamma$ represents the total decay width.

Extracting multiple complex poles poses a challenging non-convex optimization problem in a high-dimensional parameter space. To ensure robust convergence, the optimization is carried out using \texttt{MINUIT} library~\cite{James:1975} via multi-trial randomized fits. Before comparing the theoretical predictions with CLAS photoproduction measurements, the theoretical spectrum is convolved with an effective $7.0$ [MeV] Gaussian detector resolution kernel~\cite{CLAS:2013rjt} and integrated across matching energy intervals. For each of the $27$ CLAS dataset, $500$ independent optimization trials are executed with parameter sets sampled across the physical search bounds following Eq.~\eqref{eq:sampspace}.

The best fit lineshapes for the $[2_{bt}, 1_{bb}, 0_{tb}]$ pole structure model are shown in Fig.~\ref{fig:minuitfit} whereas the complete list of the extracted pole parameters across different center-of-mass energy range $W$ and charge channels are listed in Table~\ref{tab:polepars}.

The mean pole parameter values in Table~\ref{tab:poleparmean} reveals a fundamental physical distinction between the high- and low-mass singularity structures. The higher-mass pole, $\sqrt{s}_{[bt],1}^{\mathrm{(pole)}} = (1422.48 \pm 2.36) - (25.45 \pm 3.83)i$ [MeV], exists as a solitary singularity on the $[bt]$ sheet situated immediately below the $\bar{K}N$ threshold. This isolated structure is the hallmark of a dynamically generated $\bar{K}N$ molecular quasi-bound state. Its proximity to $\varepsilon_2$ provides an evidence that this higher-mass state is fundamentally a higher channel-induced hadronic molecule generated dynamically by $\bar{K}N$ interaction.

In contrast, the secondary $[bt]$ pole at $\sqrt{s}_{[bt],2}^{\mathrm{(pole)}}=(1380.32 \pm 3.46) - (85.64 \pm 13.32)i$ [MeV] and the tertiary $[bb]$ pole at $\sqrt{s}_{[bb]}^{\mathrm{(pole)}}=(1386.32 \pm 10.29) - (111.74 \pm 20.79)i$ [MeV] form a tight, numerically consistent pole-shadow pair across adjacent Riemann sheets. Their real pole masses, $[bt]: 1380.32 \pm 3.46 \,[\text{MeV}]$ and $[bb]: 1386.32 \pm 10.29$\, [\text{MeV}], agree well within $1\sigma$ standard error, while their imaginary half-decay widths, $[bt]: 85.64 \pm 13.32$ [MeV] and $[bb]: 111.74 \pm 20.79$ [MeV], similarly overlap within uncertainties. Because its pole position lies close to the open $\Sigma \pi$ threshold, it cannot form a stable bound state of the lower channel. Instead, it manifests as a broad and unstable resonance. Following Morgan's pole counting criterion~\cite{Morgan:1992}, this structure strongly indicates that the lower-mass state possesses a predominantly non-molecular nature interacting with the $\Sigma \pi$ continuum. This assertion is supported by prior analysis in \cite{Sekihara:2013wlq} which extracted an elementarity value $Z= 0.86- 0.40i$ of the lower-mass pole of the $\Lambda(1405)$ system.

Importantly, while our global CNN ensemble classification compellingly identifies the three-pole $[2_{bt}, 1_{bb}, 0_{tb}]$ structure as the dominant physical candidate, this conclusion should be understood under the consideration of the limited resolution of the current CLAS photoproduction measurements. Because shadow poles on the distant $[bb]$ sheet exert only subtle, sub-dominant distortions on physical real-energy observables, disentangling genuine tertiary poles from non-resonant effects remains challenging under existing empirical uncertainties. The necessity of a three-pole structure is therefore warranted specifically as a global, dataset-wide probabilistic consensus derived from ensemble averaging across all $27$ CLAS lineshapes, rather than an indisputable local certainty in every kinematic bin. Thus, higher-resolution and high-statistics experimental measurements is required to either decisively support or refute the presence of the shadow pole in the $[bb]$ sheet.

\section{Conclusion and Outlook}\label{sec:6}

This work presented a machine learning-based analysis of the $\Lambda(1405)$. Our findings establish that the empirical $\Sigma\pi$ invariant mass spectrum is governed by a two-state system. Crucially, our framework does not favor both the conventional single-pole hypothesis and the popular two-pole formulations. Instead, we demonstrate that the CLAS dataset requires a three-pole $[2_{bt}, 1_{bb}, 0_{tb}]$ structure. Subsequent pole parameter extraction reveal a clear physical distinction: the higher-mass pole exists as an isolated singularity on the $[bt]$ sheet below the $\bar{K}N$ threshold, indicating a dynamically generated molecular state. Conversely, the lower-mass state manifests as a pole-shadow pair across the $[bt]$ and $[bb]$ sheets just above the $\Sigma \pi$ threshold, indicative of a predominantly non-molecular state interacting with the open $\Sigma \pi$ continuum.

We have also demonstrated in this work a possible synergy between machine learning techniques and conventional fitting procedures. Specifically, a machine learning analysis of the pole structure can be implemented to constrain the interpretation space associated with the experimental data. This is then followed by a fitting procedure that employs the preferred parametrization informed by the machine learning result~\cite{Sombillo:2025izy}. Moreover, recent studies indicate that incorporating machine learning directly into the fitting procedure itself has yielded promising results, particularly in the presence of misspecified data points~\cite{Sadasivan:2025kjj}. It is worth noting that, in the fitting procedure carried out here, the same independent $\mathcal{S}$-matrix pole parametrization was used as in the generation of the training dataset. Consequently, the poles obtained from the standard fitting are not linked by coupling parameters, as is the case in other parametrizations such as the $\mathcal{K}$-matrix~\cite{Chung:1995dx}. Thus, the extracted pole positions are free from interpretative bias.

A fully unambiguous interpretation of the extracted pole structure, however, requires the attachment of a dynamical model to the obtained poles and their positions. In particular, one may turn off the channel coupling to trace the origin of the poles and thereby identify the nature of the state. Such a treatment, though, is inherently model- or theory-dependent, as the conclusions drawn will vary with the chosen dynamical framework. Thus, while our approach provides a bias‑free mapping of pole positions and highlights the synergy between machine learning and conventional fitting, the ultimate physical interpretation of these poles must be guided by carefully selected theoretical models. This underscores both the promise and the necessary caution in extending machine learning to the broader landscape of hadron spectroscopy.

\bibliography{mybib}
\appendix
\label{sec:appendix}
\setcounter{equation}{0}
\setcounter{figure}{0}
\setcounter{table}{0}
\renewcommand{\theequation}{A.\arabic{equation}}
\renewcommand{\thefigure}{A.\arabic{figure}}
\renewcommand{\thetable}{A.\arabic{table}}
\begin{table*}[!ht]
    \centering
    \setlength{\tabcolsep}{10pt}
    \renewcommand{\arraystretch}{2.0}
    \caption{\textbf{Extracted complex pole positions for the coupled-channel $\Sigma\pi-\bar{K}N$ system across energy bins.} Complete dataset listing the extracted complex pole parameters $\sqrt{s}_{\mathrm{pole}} = \mathrm{Re}(\sqrt{s}_{\mathrm{pole}}) - i\,\mathrm{Im}(\sqrt{s}_{\mathrm{pole}})$ [MeV] across nine center-of-mass energy intervals ($W$) and three decay channels ($\Sigma^+\pi^-$, $\Sigma^0\pi^0$, and $\Sigma^-\pi^+$). Columns 3 and 4 list primary and secondary pole positions on the $[bt]$ sheet, while column 5 contains pole parameter values of the tertiary $[bb]$ sheet pole. Pole parameters represented by ($\cdots$) are values that hit the artificial search grid boundaries. Because the minimizer was forced to accommodate a fixed $[2_{bt}, 1_{bb}, 0_{tb}]$ structure even in energy bins where data does not require three active poles, the algorithm likely drives the redundant poles to the grid edges to minimize their influence on the lineshape.}
    \label{tab:polepars}
    \begin{tabular}{ccccc} 
    \hline \hline
       Channel & $W$ [GeV] & $\sqrt{s}_{\,[bt],1}^{\mathrm{(pole)}}$ [MeV] & $\sqrt{s}_{\,[bt],2}^{\mathrm{(pole)}}$ [MeV] & $\sqrt{s}_{\,[bb]}^{\mathrm{(pole)}}$ [MeV]\\
    \hline
    $\Sigma^+ \pi^-$ 
    & $1.95<W<2.05$ & $1423.77 - 51.37i$ & $\cdots$ & $\cdots$ \\
    & $2.05<W<2.15$ & $1419.57 - 3.56i$ & $1379.36-60.34i$ & $1378.30 - 22.44i$\\
    & $2.15<W<2.25$ & $1423.74 - 28.85i$ & $1368.73 - 1.70i$ & $1354.11 - 12.01i$\\
    & $2.25<W<2.35$ & $1432.86 - 8.91 i$ & $1419.01 - 215.41i$ & $1435.88 - 150.36i$\\
    & $2.35<W<2.45$ & $1410.02 - 7.52i$ & $1378.46 - 37.23i$ & $\cdots$\\
    & $2.45<W<2.55$ & $1438.04 - 8.93i$ & $1395.65 - 94.69i$ & $\cdots$\\
    & $2.55<W<2.65$ & $1412.00 - 56.49i$ & $\cdots$ & $\cdots$\\
    & $2.65<W<2.75$ & $1417.13 - 43.07i$ & $\cdots$ & $\cdots$\\
    & $2.75<W<2.85$ & $1421.47 - 14.18i$ & $1378.40 - 86.31i$ & $1398.17 - 84.09i$\\
    \hline
    $\Sigma^0 \,\pi^0$ 
    & $1.95<W<2.05$ & $1414.04 - 17.39i$ & $1340.89 - 46.52i$ & $\cdots$\\
    & $2.05<W<2.15$ & $1418.04 - 8.22i$ & $\cdots$ & $\cdots$\\
    & $2.15<W<2.25$ & $1391.10 -11.70i$ & $1389.16 - 54.04i$ & $\cdots$\\
    & $2.25<W<2.35$ & $1430.81 - 19.61i$ & $1388.51 - 39.99i$ & $\cdots$\\
    & $2.35<W<2.45$ & $1414.90 - 36.59i$ & $1395.59 - 6.67i$ & $\cdots$\\
    & $2.45<W<2.55$ & $1426.09 - 15.62i$ & $1392.18 - 112.16i$ & $1429.74 - 39.26i$\\
    & $2.55<W<2.65$ & $1431.05 - 9.20i$ & $1388.31 - 55.91i$ & $1339.32 - 33.19i$\\
    & $2.65<W<2.75$ & $1433.90 - 9.52i$ & $1397.05 - 50.26i$ & $1331.36 - 92.82i$\\
    & $2.75<W<2.85$ & $1433.66 - 3.01i$ & $1403.19 - 73.93i$ & $1431.41 - 24.32i$\\
    \hline
    $\Sigma^- \pi^+$ 
    & $1.95<W<2.05$ & $1433.87 - 21.34i$ & $1364.64 - 126.89i$ & $1333.19 - 178.83i$\\
    & $2.05<W<2.15$ & $1442.91 - 67.09i$ & $1380.48 - 272.55i$ & $\cdots$\\
    & $2.15<W<2.25$ & $1389.66 - 75.91i$ & $1366.77 - 1.28i$ & $\cdots$\\
    & $2.25<W<2.35$ & $1424.60 - 27.65i$ & $1366.86 - 94.36i$ & $1367.05 - 122.08i$\\
    & $2.35<W<2.45$ & $1426.60 - 34.92i$ & $1365.35 - 121.16i$ & $1348.33 - 259.94i$\\
    & $2.45<W<2.55$ & $1419.45 - 19.36i$ & $1372.36 - 113.92i$ & $1382.79 - 111.82i$\\
    & $2.55<W<2.65$ & $1423.46 - 47.55i$ & $1377.60 - 63.39i$ & $1424.80 - 184.70i$\\
    & $2.65<W<2.75$ & $1425.24 - 14.26i$ & $1362.19 - 130.03i$ & $1398.21 - 109.54i$\\
    & $2.75<W<2.85$ & $1429.02 - 25.33i$ & $1376.72 - 110.92i$ & $1442.17 - 250.67i$\\
    \hline \hline
    \end{tabular}
\end{table*}

\end{document}